\documentstyle[aps,12pt]{revtex}
\begin{document}
\title{Parity violating elastic electron scattering and neutron
density distributions in the Relativistic Hartree-Bogoliubov model}
\bigskip
\author{D. Vretenar$^{1,2,3}$, P. Finelli$^{2}$, A. Ventura$^{4}$,
G.A. Lalazissis$^{1,5}$, and P. Ring$^{1}$}
\bigskip
\address{
$^{1}$ Physik-Department der Technischen Universit\"at M\"unchen,
D-85748 Garching, Germany\\
$^{2}$ Physics Department, University of Bologna, and INFN - Bologna, 
I-40126 Bologna, Italy\\
$^{3}$ Physics Department, Faculty of Science, University of
Zagreb, 10 000 Zagreb, Croatia\\
$^{4}$Centro Dati Nucleari, ENEA, Via Martiri di Monte Sole 4,
I-40129 Bologna, Italy\\
$^{5}$ Wright Nuclear Structure Laboratory, Yale University, 
New Haven, Connecticut 06520\\}
\maketitle
\bigskip
\bigskip
\begin{abstract}
Parity violating elastic electron scattering
on neutron-rich nuclei is described in the framework
of relativistic mean-field theory. Self-consistent 
ground state density distributions of Ne, Na, Ni and Sn
isotopes are calculated with the relativistic Hartree-
Bogoliubov model, and the resulting neutron radii are 
compared with available experimental data. For the elastic
scattering of 850 MeV electrons on these nuclei, the 
parity-violating asymmetry parameters are calculated 
using a relativistic optical model with inclusion of 
Coulomb distortion effects. The asymmetry parameters
for chains of isotopes are compared, and their relation
to the Fourier transforms of neutron densities is studied.
It is shown that parity-violating asymmetries are 
sensitive not only to the formation of the neutron skin,
but also to the shell effects of the neutron density
distribution.
\end{abstract}
\vspace{1 cm}
{PACS number(s):} {21.10.Gv, 21.60.-n, 25.30.Bf, 27.30.+t, 
27.50.+e, 27.60.+j}\\
\vspace{1 cm}\\
\section{Introduction}
Measurements of ground state distributions of nucleons provide
fundamental nuclear structure informations. The ground state 
densities reflect the basic properties of effective nuclear 
forces, and an accurate description of these distributions presents 
the primary goal of nuclear structure models.

Extremely accurate data on charge densities, and therefore on proton 
distributions in nuclei, are obtained from elastic scattering of electrons. 
Experimental data of comparable precision on neutron density distributions
are, however, not available. It is much more difficult to measure the 
distribution of neutrons, though more recently accurate data on differences
in radii of the neutron and proton density distributions have been 
obtained. Various experimental methods have been used, or suggested, 
for the determination of the neutron density in nuclei~\cite{Bat.89}.
Among them, one that is also very interesting from the theoretical
point of view, is parity violating elastic electron scattering. 
In principle, the elastic scattering of longitudinally polarized
electrons provides a direct and very accurate measurement of the 
neutron distribution. It has been shown that the parity-violating
asymmetry parameter, defined as the difference between cross sections
for the scattering of right- and left-handed longitudinally
polarized electrons, produces direct information on the 
Fourier transform of the neutron density~\cite{DDS.89}.

In a recent article~\cite{Hor.98}, Horowitz has used a relativistic 
optical model to calculate the parity-violating asymmetry parameters
for the elastic scattering of 850 MeV electrons on a number of spherical,
doubly closed-shell nuclei. Ground state densities were calculated 
using a three-parameter Fermi formula and the relativistic mean-field
model. Coulomb distortion corrections to the parity-violating asymmetry
were calculated exactly. It has been shown that a parity violation 
experiment to measure the neutron density in a heavy nucleus is feasible.

Informations about the distribution of neutrons in nuclei should also 
constrain the isovector channel of the nuclear matter energy functional.
A correct parameterization of the isovector channel of effective nuclear 
forces is essential for the description of unique phenomena in exotic
nuclei with extreme isospin values. For neutron-rich nuclei, these 
include the occurrence of nuclei with diffuse neutron densities, the 
formation of the neutron skin and the neutron halo. At the neutron drip-line
different proton and neutron quadrupole deformations are expected, and
these will give rise to low-energy isovector modes. 

In the present work we describe parity violating elastic electron scattering
on neutron-rich nuclei in the framework of relativistic mean-field theory.
The ground state density distributions will be calculated with 
the  relativistic Hartree-Bogoliubov (RHB) model. This model 
represents a relativistic extension of the Hartree-Fock-Bogoliubov
(HFB) framework, and it has been successfully applied in studies of the
neutron halo phenomenon in light nuclei \cite{PVL.97}, 
properties of light nuclei near the neutron-drip line \cite{LVP.98},
ground state properties of Ni and Sn isotopes \cite{LVR.98},
the deformation and shape coexistence phenomena that result from the
suppression of the spherical N=28 shell gap in neutron-rich nuclei
\cite{LVR.98a}, the structure of proton-rich nuclei and the 
phenomenon of ground state proton emission \cite{LVR.99,VLR.99,LVR.99a}.
In particular, it has been shown that neutron radii, calculated 
with the RHB model, are in excellent agreement with experimental
data \cite{LVP.98,LVR.98}. In studies of the structure of nuclei
far from the $\beta$-stability line, both on the proton- and
neutron-rich sides, it is important to use models that include a 
unified description of mean-field and pairing correlations~\cite{DNW.96}. 
For the ground state density distributions of neutron-rich 
nuclei in particular, the neutron Fermi level is found close to the 
particle continuum and the lowest particle-particle modes are 
embedded in the continuum. An accurate description of neutron densities 
can only be obtained with a correct treatment of the scattering 
of neutron pairs from bound states into the positive energy 
continuum.

Starting from the RHB self-consistent ground state neutron densities
in isotope chains that also include neutron-rich nuclei, we will calculate 
the parity-violating asymmetry parameters for the elastic scattering 
of 850 MeV electrons. The main point in the present analysis will be
to determine how sensitive the asymmetry parameters are to the variations
of the neutron density distribution along an isotope chain. An interesting
question is whether parity violating electron scattering could be used, in 
principle at least, to measure the neutron skin, or even the formation of 
the neutron halo. Of course, studies of electron scattering from exotic nuclei
require very complex experimental facilities, as for example 
the double storage ring MUSES, under construction at RIKEN. 
By injecting an electron beam, generated by a linear accelerator, in one of the
storage rings, and storing a beam of unstable nuclei in the other, 
collision experiments between radioactive beams and electrons
could be performed.

In Section II we present an outline of the relativistic Hartree-Bogoliubov
model and calculate the self-consistent ground state neutron densities 
of Ne, Na, Ni and Sn isotopes. For the elastic scattering of longitudinally 
polarized electrons on these nuclei, in section III we use a relativistic    
optical model to calculate the parity-violating asymmetry parameters. 
Coulomb distortion corrections are included in the calculation, and the
resulting asymmetries are related to the Fourier transforms of the 
neutron densities. The results are summarized in Section IV.

\section{Relativistic Hartree-Bogoliubov description of ground state 
densities}

In the framework of the relativistic Hartree-Bogoliubov 
model~\cite{LVR.99,Rin.96}, the ground state of a nucleus $\vert \Phi >$ 
is represented by the product of independent
single-quasiparticle states. These states are eigenvectors of the
generalized single-nucleon Hamiltonian which
contains two average potentials: the self-consistent mean-field
$\hat\Gamma$ which encloses all the long range particle-hole ({\it ph})
correlations, and a pairing field $\hat\Delta$ which sums
up the particle-particle ({\it pp}) correlations. The
single-quasiparticle equations result from the variation of the
energy functional with respect to the hermitian density matrix $\rho$
and the antisymmetric pairing tensor $\kappa$. The relativistic
Hartree-Bogoliubov equations read
\begin{eqnarray}
\label{equ.2.2}
\left( \matrix{ \hat h_D -m- \lambda & \hat\Delta \cr
		-\hat\Delta^* & -\hat h_D + m +\lambda} \right)
		\left( \matrix{ U_k({\bf r}) \cr V_k({\bf r}) } \right) =
		E_k\left( \matrix{ U_k({\bf r}) \cr V_k({\bf r}) } \right).
\end{eqnarray}
$\hat h_D$ is the single-nucleon Dirac Hamiltonian, $m$ is the nucleon 
mass, and $\hat\Delta$ denotes the pairing field. The column vectors 
represent the quasi-particle spinors and $E_k$ are the quasi-particle
energies. The chemical potential $\lambda$ has to be determined by
the particle number subsidiary condition in order that the
expectation value of the particle number operator
in the ground state equals the number of nucleons.

The Hamiltonian $\hat h_D$ describes the dynamics of the 
relativistic mean-field model~\cite{SW.97}:
nucleons are described as point particles;
the theory is fully Lorentz invariant; the nucleons move independently in
the mean fields which originate from the nucleon-nucleon interaction.
Conditions of causality and Lorentz invariance impose that the
interaction is mediated by the exchange of point-like 
effective mesons, which couple to the nucleons at local vertices:
the isoscalar scalar $\sigma$-meson, the isoscalar vector
$\omega$-meson, and the isovector vector $\rho$-meson.
The single-nucleon Hamiltonian in the Hartree approximation
reads
\begin{equation}
\hat{h}_{D}=-i{\mathbf{\alpha \cdot \nabla }}+\beta (m+g_{\sigma }\sigma
({\mathbf r}))+g_{\omega }\omega ({\mathbf r})+
g_{\rho }\tau_{3}\rho_{3} ({\mathbf r})+
e{\frac{{(1-\tau _{3})}}{2}}A({\mathbf r}),  \label{dirh}
\end{equation}
where $\sigma$, $\omega$, and $\rho$ denote the mean-field meson 
potentials. $g_{\sigma }$, $g_{\omega }$ and $g_{\rho }$ are the 
corresponding meson-nucleon coupling constants, and the
photon field $A$ accounts for the electromagnetic
interaction. The meson potentials are determined self-consistently
by the solutions of the corresponding Klein-Gordon equations. 
The source terms for these equations are sums of bilinear products 
of baryon amplitudes, calculated in the {\it no-sea} approximation.

The pairing field $\hat\Delta $ in the RHB single-quasiparticle 
equations (\ref{equ.2.2}) is defined  
\begin{equation}
\label{equ.2.5}
\Delta_{ab} ({\bf r}, {\bf r}') = {1\over 2}\sum\limits_{c,d}
V_{abcd}({\bf r},{\bf r}') {\bf\kappa}_{cd}({\bf r},{\bf r}'),
\end{equation}
where $a,b,c,d$ denote quantum numbers
that specify the Dirac indices of the spinors,
$V_{abcd}({\bf r},{\bf r}')$ are matrix elements of a
general two-body pairing interaction, and the pairing
tensor is 
\begin{equation}
{\bf\kappa}_{cd}({\bf r},{\bf r}') =
\sum_{E_k>0} U_{ck}^*({\bf r})V_{dk}({\bf r}').
\end{equation}

The input parameters of the RHB model are the coupling constants and the
masses for the effective mean-field Hamiltonian, and the effective
interaction in the pairing channel. In most applications we have
used the NL3 set of meson masses and meson-nucleon coupling
constants~\cite{LKR.97} for the effective interaction
in the particle-hole channel: $m=939$ MeV, $m_{\sigma}=508.194$ MeV,
$m_{\omega}=782.501$ MeV, $m_{\rho}=763.0$ MeV,
$g_{\sigma}=10.217$, $g_2=-10.431$ fm$^{-1}$, $g_3=-28.885$,
$g_{\omega}=12.868$ and  $g_{\rho}=4.474$.
Results of NL3 model calculations have been found in excellent 
agreement with experimental data, both for stable nuclei 
and for nuclei far away from the line of $\beta $-stability.
The NL3 interaction will also be used in the present analysis
of ground state neutron densities. For the pairing field we employ 
the pairing part of the Gogny interaction
\begin{equation}
V^{pp}(1,2)~=~\sum_{i=1,2}
e^{-(( {\bf r}_1- {\bf r}_2)
/ {\mu_i} )^2}\,
(W_i~+~B_i P^\sigma
-H_i P^\tau -
M_i P^\sigma P^\tau),
\end{equation}
with the set D1S \cite{BGG.84} for the parameters
$\mu_i$, $W_i$, $B_i$, $H_i$ and $M_i$ $(i=1,2)$.
This force has been carefully adjusted to the pairing
properties of finite nuclei all over the periodic table.
In particular, the finite range of the Gogny force
automatically guarantees a proper cut-off in momentum space.
On the phenomenological level, the fact that we are using 
a non-relativistic interaction in the pairing channel of 
a relativistic Hartree-Bogoliubov model, has no influence
on the calculated ground state properties. A detailed discussion
of this approximation can be found, for instance, 
in Ref.~\cite{LVR.99}. 

The ground state of a nucleus results from a self-consistent solution
of the RHB single-quasiparticle equations (\ref{equ.2.2}).
The iteration procedure is performed in the quasi-particle basis.
A simple blocking prescription is used in the calculation 
of odd-proton and/or odd-neutron systems.
The resulting eigenspectrum is
transformed into the canonical basis of single-particle
states, in which the RHB ground state takes the
BCS form. The transformation determines the energies
and occupation probabilities of the canonical states.

In Ref.~\cite{LVR.98} we have performed a detailed analysis 
of ground state properties of Ni ($28\leq N\leq 50$) and
Sn ($50\leq N\leq 82$) nuclei in the framework of the
RHB model. In a comparison with available experimental
data, we have shown that the NL3 + Gogny D1S effective
interaction provides an excellent description of binding energies,
neutron separation energies, and proton and neutron $rms$ radii,
both for even and odd-A isotopes. The RHB model predicts
a reduction of the spin-orbit potential with the
increase of the number of neutrons. The resulting
energy splittings between spin-orbit partners have been 
discussed, as well as pairing properties calculated
with the finite range effective interaction in the $pp$ channel.
In Figs. \ref{figA} and \ref{figB} we plot the self-consistent
ground state neutron densities for the even-A Ni ($30\leq N\leq 48$)
and Sn ($56\leq N\leq 74$) isotopes. The density profiles display
pronounced shell effects and a gradual increase of neutron radii.
In the inserts we include the corresponding differences between
neutron and proton $rms$ radii. For Ni, the value of $r_n - r_p$ 
increases from $\approx 0.03$ fm for $^{58}$Ni to 0.39 fm for $^{76}$Ni.
The neutron skin is less pronounced for the Sn isotopes: 
$r_n - r_p = 0.27$ fm for $^{124}$Sn. In Fig. \ref{figC} 
the difference of the neutron and proton $rms$ radii for Sn isotopes 
are compared with recent experimental data ~\cite{Kra.99}. The experimental
values result from measured cross sections of the isovector spin-dipole
resonances in Sn nuclei. The agreement between theoretical and 
experimental values is very good. The calculated differences are 
slightly larger than the measured values, but they are still within 
the experimental error bars. This result, together with the analysis 
of Ref.~\cite{LVR.98}, indicates that the ground state neutron 
densities of the Ni and Sn isotope chains are correctly described
by the RHB model with the NL3 + Gogny D1S effective interaction.

The first experimental evidence of the formation of the neutron 
skin along a chain of stable and unstable isotopes was reported 
in Ref.~\cite{Suz.95} for the Na nuclei. By combining $rms$ charge
radii, determined from isotope-shift data, with interaction cross 
sections of Na isotopes on a carbon target, it has been shown that
the neutron skin monotonically increases to 0.4 fm in neutron-rich
$\beta$-unstable Na nuclei. In Ref.~\cite{LVP.98} we have applied the 
RHB model in the description of properties of light nuclei with large
neutron excess. The ground state properties of a number  
of neutron-rich nuclei have been analyzed: the location 
of the neutron drip-line, the reduction of the spin-orbit interaction,
$rms$ radii, changes in surface properties, and the formation of the
neutron skin and the neutron halo. In particular, we have also calculated
the chain of Na isotopes. The neutron density profiles are plotted
in Fig. \ref{figD}, with the differences of the neutron and proton $rms$ 
radii in the insert. The calculations have been performed assuming 
spherical symmetry, i.e. deformations of proton and neutron densities
were not taken into account. The blocking procedure has been used both
for protons and neutrons. Strong shell effects are observed in the 
interior. In the central region the neutron density increases from 
$\approx 0.065$ fm$^{-3}$ for $^{26}$Na, to more than $0.09$ fm$^{-3}$ 
for $^{27}$Na. This effect, of course, corresponds to the filling of  
the $s{1\over 2}$ neutron orbital. At $r \approx 2$ fm the density 
decreases from  $0.1$ fm$^{-3}$ to  $0.09$ fm$^{-3}$. 
The calculated neutron radii are compared with the experimental
values ~\cite{Suz.95} in Fig. \ref{figE}. The model reproduces the trend  
of the experimental data and, with a possible exceptions of $N = 11$,
the theoretical values are in excellent agreement with the measured
radii. The smooth increase of $r_n - r_p$ after $N = 11$ (Fig. \ref{figD})
is interpreted as the formation of the neutron skin. For $N = 11$
the number of protons and neutrons is the same, and therefore
both the odd proton and odd neutron occupy the $s-d$
orbitals with the same probabilities. An explicit proton-neutron
short-range interaction, not included in our model, is probably 
responsible for a possible reduction of the neutron radius in this
nucleus.

In some loosely bound systems at the drip-lines,
the neutron density distribution displays an extremely long
tail: the neutron halo. The resulting large interaction
cross sections have provided the first experimental evidence
for halo nuclei~\cite{Tani.85}. The neutron halo phenomenon
has been studied with a variety of theoretical 
models~\cite{Tani.95,HJJ.95,MR.96}.
For very light nuclei in particular, models based on the
separation into core plus valence space nucleons
(three-body Borromean systems) have been employed.
In Ref. ~\cite{PVL.97} the RHB model has been
applied in the study of the formation of the neutron halo
in the mass region above the s-d shell. Model calculations
with the NL3 + Gogny D1S effective interaction predict the 
occurrence of neutron halo in heavier Ne isotopes. The study
has shown that, in a mean-field description, the neutron halo and
the stability against nucleon emission can only be explained
with the inclusion of pairing correlations.
Both the properties of single-particle states near the
neutron Fermi level, and the pairing interaction, are important
for the formation of the neutron halo. In Fig. \ref{figF} we
plot the proton and neutron density distributions for $^{30}$Ne,
$^{32}$Ne and $^{34}$Ne. The proton density profiles do not change 
with the number of neutrons. The neutron density distributions
display an abrupt change between $^{30}$Ne and $^{32}$Ne. A long tail
emerges, revealing the formation of a multi-particle halo.
The formation of the neutron halo is related to the quasi-degeneracy
of the triplet of states 1f$_{7/2}$, 2p$_{3/2}$ and  2p$_{1/2}$.
For $N \leq 22$ the triplet of states is in the continuum, it 
approaches zero energy at $N=22$, and a gap is formed between these 
states and all other states in the continuum. The pairing interaction
promotes neutrons from the 1f$_{7/2}$ orbital to the 2p
levels. Since these levels are so close in energy, the
total binding energy does not change significantly. Due to
their small centrifugal barrier, the 2p$_{3/2}$ and
2p$_{1/2}$ neutron orbitals form the halo.

\section{Parity violating elastic electron scattering}

In this section we will illustrate how the neutron density
distributions shown in Figs. \ref{figA} - \ref{figF} can be 
measured with polarized elastic electron scattering. 
We will calculate the parity-violating asymmetry parameter,
defined as the difference between cross sections for the scattering
of right- and left-handed longitudinally polarized electrons. 
This difference arises from the interference of one-photon and 
$Z^0$ exchange. As it has been shown in Ref. \cite{DDS.89}, 
the asymmetry in the parity violating elastic polarized electron 
scattering represents an almost direct measurement of the 
Fourier transform of the neutron density.

The calculation procedure closely follows the derivation and 
definitions of Ref. \cite{Hor.98}. We consider the elastic
electron scattering on a spin-zero nucleus, i.e. on the potential
\begin{equation}
\hat V(r) = V(r) + \gamma_5 A(r),
\label{pot1}
\end{equation}
where V(r) is the Coulomb potential, and A(r) results from the 
weak neutral current amplitude 
\begin{equation}
A(r)= {G_F\over 2^{3/2}} \rho_W(r).
\label{pot2}
\end{equation}
The weak charge density is defined 
\begin{equation}
\rho_W(r)=\int d^3r^\prime G_E(|{\bf r}-{\bf r}^\prime|)
[-\rho_n(r^\prime) + (1-4{\rm sin}^2\Theta_W)\rho_p(r^\prime)],
\label{rhoW}
\end{equation}
where $\rho_n$ and $\rho_p$ are point neutron and proton densities and the
electric form factor of the proton is $G_E(r)\approx {\Lambda^3 
\over 8\pi}e^{-\Lambda r}$ with $\Lambda=4.27$ fm$^{-1}$.
sin$^2\Theta_W=0.23$ for the Weinberg angle.

In the limit of vanishing electron mass, the electron spinor $\Psi$
defines the helicity states 
\begin{equation}
\Psi_\pm={1\over 2}(1\pm\gamma_5)\Psi,
\end{equation}
which satisfy the Dirac equation 
\begin{equation}
[{\bf \alpha}\cdot {\bf p} + V_\pm(r)]\Psi_\pm = E \Psi_\pm,
\end{equation}
with 
\begin{equation}
V_\pm(r) = V(r) \pm A(r).
\end{equation}
The parity-violating asymmetry $A_l$, or helicity asymmetry, is 
defined 
\begin{equation}
A_l={d\sigma_+/d\Omega - d\sigma_-/d\Omega \over
       d\sigma_+/d\Omega + d\sigma_-/d\Omega},
\label{AP}
\end{equation}
where $+(-)$ refers to the elastic scattering on the potential $V_\pm(r)$.

The calculation starts with the self-consistent relativistic 
Hartree-Bogoliubov ground state proton and neutron densities.
The charge and weak densities are calculated by folding 
the point proton and neutron densities (see Eq. (\ref{rhoW})).
The resulting Coulomb potential $V(r)$ and weak potential 
$A(r)$ (\ref{pot2}) are used to construct $V_\pm(r)$.
The cross sections for elastic electron scattering are 
obtained by summing up the phase shifts which result from the 
numerical solution of the partial wave Dirac equation.
The calculation includes the Coulomb distortion effects.
The cross sections for positive and negative helicity electron 
states are calculated, and the resulting asymmetry parameter 
$A_l$ is plotted as a function of the scattering angle $\theta$, or 
the momentum transfer $q$. 

In order to check the correctness and accuracy of the computer 
code which calculates the 
elastic electron scattering cross sections, we have performed the
same tests as those reported in Ref.~\cite{Hor.98}: experimental 
data are reproduced for the
elastic cross sections from $^{208}$Pb at 502 MeV~\cite{Fro.77}; 
plane wave approximation results are reproduced; it is verified that 
the asymmetry parameter $A_l$ is linear in the potential $A(r)$ 
(\ref{pot2}). We have also reproduced the parity-violating 
asymmetries calculated in Ref.~\cite{Hor.98}: 
elastic scattering at 850 MeV on $^{16}$O,  $^{48}$Ca, and 
$^{208}$Pb (relativistic mean-field densities), as well as
on the three-parameter Fermi densities.

In Figs. \ref{figG} and \ref{figH} we plot the parity-violating 
asymmetry parameters $A_l$ for the $^{58-76}$Ni isotopes, for 
the elastic electron scattering at 500 MeV and 850 MeV, respectively.
The ground state neutron densities for these nuclei are shown in 
Fig. \ref{figA}. For electron energies below 500 MeV the asymmetry
parameters are small ($< 10^{-5}$), and the differences between 
neighboring isotopes are $< 10^{-6}$. At 850 MeV (the energy for 
which most of the calculations of Ref.~\cite{Hor.98} have been 
performed), the values of $A_l$ are of the order of $ 10^{-5}$. 
The differences between neighboring isotopes ($\approx 10^{-6}$) 
are especially pronounced for  $^{58-66}$Ni, at $\theta \approx 
20^o$ and $\theta \approx 35^o$. These differences reflect the
strong shell effects calculated for the ground state neutron 
densities of the lighter Ni isotopes (see Fig. \ref{figA}). 
The heavier Ni isotopes display more uniform neutron densities
in the interior, and the resulting asymmetry parameters $A_l$
are not very different. Similar results are also obtained 
at 1000 MeV electron energy. Of course, above 1 GeV the 
approximation of elastic scattering on continuous charge and 
weak densities is not valid any more, and the structure of 
individual nucleons becomes important. In the remainder of 
this section we show the results for $A_l$ at 850 MeV
electron energy. For all chains of isotopes we have also
calculated the asymmetry parameters at 250 MeV, 500 MeV and
1000 MeV. The energy dependence, however, is similar 
to that observed for the Ni isotopes: below 850 MeV the 
differences in the calculated $A_l$ for the neighboring isotopes
are too small. 

The figures of merit
\begin{equation}
F=A_l^2{{d\sigma}\over {d\Omega}},
\label{fom}
\end{equation}
for 850 MeV electron scattering on the Ni isotopes are shown in 
Fig. \ref{figI}. The figure of merit is, of course, strongly
peaked at forward angles (see also Fig.12 of Ref.~\cite{Hor.98}).
In order to emphasize the differences between Ni isotopes, 
the $F$'s are plotted in the interval $10^o \leq \theta \leq 30^o$. 
The figure of merit defines the optimal kinematics for
an experiment in which the neutron density distribution could
be determined from the measured parity-violating asymmetries.

The asymmetry parameter $A_l$ provides a direct measurement
of the Fourier transform of the neutron density~\cite{DDS.89}.
In Figs. \ref{figJ} and \ref{figK} we plot the asymmetries
for the Ni isotopes (Fig. \ref{figH}) as functions of the 
momentum transfer $q = 2 E sin{\theta /2}$, and compare
them with the squares of the Fourier transforms of the neutron 
densities 
\begin{equation}
F(q) = {4\pi \over q} \int dr~r^2 j_0(qr) \rho_n(r).
\end{equation}
The differences between the 
asymmetries can be directly related to the form factors. 
Note that the positions of the minima of $A_l$ correspond 
almost exactly to the minima of the form factors. Of course, 
the agreement would have been perfect if we had plotted
the Fourier transforms of the weak density (\ref{rhoW}), but 
the differences are indeed very small. More important is the 
observation that the measurement of the parity-violating asymmetry
at high momentum transfer might provide information about the 
details of the density profile of the neutron distribution. 

In Fig. \ref{figC} we have shown that the differences in the 
neutron and proton $rms$ radii, calculated with the RHB 
NL3 + Gogny D1S model, are in very good agreement with 
recent experimental data on Sn isotopes. In Fig. \ref{figL}
we plot the asymmetry parameters $A_l$ for the 850 MeV 
elastic electron scattering on the even $^{106-124}$Sn isotopes.
The angular dependence is similar to that observed for 
the Ni nuclei; significant differences between neighboring 
isotopes are only found at $\theta > 20^o$. The asymmetry
parameters are compared to the Fourier transforms of the
Sn neutron densities in Figs. \ref{figM} and \ref{figN}.

The Na isotopes (Figs. \ref{figD} and \ref{figE}) display the 
formation of the neutron skin, as well as strong shell effects
in the central region of neutron densities. These shell effects
are clearly seen in the plots of the asymmetry parameters $A_l$
in Fig. \ref{figO}. Especially pronounced is the transition 
between $^{26}$Na and $^{27}$Na, which corresponds to the 
filling of the $s{1\over 2}$ neutron orbital. The Fourier
transforms of neutron densities are compared to the 
asymmetry parameters in Figs. \ref{figP} and \ref{figQ}.
The differences between the minima of the asymmetry parameters
for neighboring isotopes are the same as the differences
between the minima of the Fourier transforms of the densities.
In principle, it should be possible to deduce the neutron 
density distribution from the measured asymmetries.

An interesting theoretical question is whether the asymmetry 
parameters are sensitive to the formation of the neutron halo, 
i.e. whether parity violating electron scattering could be 
used to detect the formation of the halo. For the even Ne nuclei,
the neutron halo phenomenon is illustrated in Fig. \ref{figF}.
The tail in the neutron density develops in $^{32}$Ne. The 
Fourier transforms of neutron densities and the asymmetry
parameters $A_l$ are shown in Fig. \ref{figR}. At small momentum
transfer the differences in the asymmetry parameters are 
very small. Only at $q \approx 2.5$ fm$^{-1}$ the differences
are of the order of $10^{-5}$. This probably means that, even 
if it became possible to measure polarized elastic electron 
scattering on extremely neutron-rich nuclei, the parity-violating 
asymmetries would not be sensitive to the formation
of the neutron halo.

\section{Conclusions}

The relativistic mean-field theory has been used to study the 
parity violating elastic electron scattering on neutron-rich nuclei.
The parity-violating asymmetry parameter, defined as the difference
between cross sections for the scattering of right- and 
left-handed longitudinally polarized electrons, provides 
direct information about the neutron density distribution.

The ground state neutron densities of neutron-rich Ne, Na, 
Ni and Sn have been calculated with the relativistic
Hartree-Bogoliubov model. 
The NL3 effective interaction has been used for the
mean-field Lagrangian, and pairing correlations have been described
by the pairing part of the finite range Gogny interaction D1S.
The NL3 + Gogny D1S interaction produces results in excellent
agreement with experimental data, not only for spherical
and deformed $\beta$-stable nuclei, but also for nuclear
systems with large isospin values on both sides of the valley
of $\beta$-stability. In the present work, in particular, 
the calculated neutron $rms$ radii are shown to reproduce
recent experimental data on Na and Sn isotopes.

Starting from the relativistic Hartree-Bogoliubov solutions for 
the self-consistent ground states, the charge and weak
densities are calculated by folding the point proton and
neutron densities. These densities define the Coulomb and
weak potentials in the Dirac equation for the massless electron.
The partial wave Dirac equation is solved with the inclusion
of Coulomb distortion effects, and the cross sections for 
positive and negative helicity electron states are calculated.
The parity-violating asymmetry parameters are plotted as
functions of the scattering angle $\theta$, or the
momentum transfer $q$, and they are compared with the 
Fourier transforms of the neutron density distributions. 

We have compared the parity-violating asymmetry parameters
for chains of neutron-rich isotopes. For low electron
energies ($\leq 500$ MeV), the differences between 
neighboring isotopes are very small. At 850 MeV, 
significant differences in the parity-violating 
asymmetries are found at $\theta > 20^o$. They can be 
related to the differences in the neutron density 
distributions. In particular, if plotted as function
of the momentum transfer $q$, the asymmetry parameter
can be related to the Fourier transform of the 
neutron density. It has been shown that the 
parity violating elastic electron scattering
is sensitive to the formation of the neutron skin
in Na, Ni and Sn isotopes, and also to the 
shell effects of the neutron density distributions.
On the other hand, from the example of neutron-rich
Ne nuclei, it appears that the asymmetry parameters
would not be sensitive to the formation of the neutron
halo. We conclude that, if it became possible to measure
parity violating elastic electron scattering
on neutron-rich nuclei, the asymmetry parameters
would provide detailed information on neutron
density distributions, neutron radii, 
and differences between neutron and charge radii.
This knowledge is, of course, essential for constraining the 
isovector channel of effective interactions in nuclei, 
and therefore for our understanding of the structure of
nuclear systems far from the $\beta$-stability line.

\bigskip
\begin{center}
{\bf ACKNOWLEDGMENTS}
\end{center}

This work has been supported in part by the
Bundesministerium f\"ur Bildung und Forschung under
project 06 TM 875, and by the Deutsche Forschungsgemeinschaft.
The relativistic optical code
is based on the program for elastic electron scattering 
DREPHA, written by B. Dreher, J. Friedrich and S. Klein.

\newpage
{\bf Figure Captions}
\bigskip

\begin{figure}
\caption{Self-consistent RHB single-neutron density distributions
for even-N Ni ($30 \leq N\leq 48$) nuclei, calculated with 
the NL3 + Gogny D1S effective interaction. The differences between
neutron and proton $rms$ radii are shown in the insert.}
\label{figA}
\end{figure}

\begin{figure}
\caption{Same as in Fig. \ref{figA}, but for Sn 
isotopes ($56 \leq N\leq 74$).}
\label{figB}
\end{figure}

\begin{figure}
\caption{Calculated differences between neutron and proton
$rms$ radii in Sn isotopes (squares), compared with 
experimental data from Ref.~[13].}
\label{figC}
\end{figure}

\begin{figure}
\caption{Self-consistent RHB single-neutron density distributions
for Na ($12 \leq N\leq 21$) isotopes.}
\label{figD}
\end{figure}

\begin{figure}
\caption{Comparison of the calculated neutron $rms$ radii for Na
isotopes with experimental data from Ref.~[14].}
\label{figE}
\end{figure}

\begin{figure}
\caption{Self-consistent RHB proton and neutron densities 
for $^{30,32,34}$Ne.}
\label{figF}
\end{figure}

\begin{figure}
\caption{Parity-violating asymmetry parameters $A_l$ for 
elastic scattering from $^{58-76}$Ni at 500 MeV, 
as functions of the scattering angle $\theta$.}
\label{figG}
\end{figure}

\begin{figure}
\caption{Parity-violating asymmetry parameters $A_l$ for 
elastic scattering from $^{58-76}$Ni at 850 MeV, 
as functions of the scattering angle $\theta$.}
\label{figH}
\end{figure}

\begin{figure}
\caption{Figures of merit (\ref{fom}) as functions of
the scattering angle $\theta$, for elastic scattering from
$^{58-76}$Ni at 850 MeV.}
\label{figI}
\end{figure}

\begin{figure}
\caption{Parity-violating asymmetry parameters $A_l$
(upper panel) and squares of normalized Fourier transforms
of neutron densities (lower panel), as functions of the 
momentum transfer $q$, for elastic scattering from
$^{58-66}$Ni at 850 MeV.}
\label{figJ}
\end{figure}

\begin{figure}
\caption{Same as in Fig. \ref{figJ}, but for 
elastic scattering from $^{68-76}$Ni at 850 MeV.}
\label{figK}
\end{figure}

\begin{figure}
\caption{Parity-violating asymmetry parameters $A_l$ for
elastic scattering from $^{106-124}$Sn at 850 MeV,
as functions of the scattering angle $\theta$.}
\label{figL}
\end{figure}

\begin{figure}
\caption{Parity-violating asymmetry parameters $A_l$
(upper panel) and squares of normalized Fourier transforms
of neutron densities (lower panel), as functions of the
momentum transfer $q$, for elastic scattering from
$^{106-114}$Sn at 850 MeV.}
\label{figM}
\end{figure}

\begin{figure}
\caption{Same as in Fig. \ref{figM}, but for 
elastic scattering from $^{116-124}$Sn at 850 MeV.}
\label{figN}
\end{figure}

\begin{figure}
\caption{Parity-violating asymmetry parameters $A_l$ for
elastic scattering from $^{23-32}$Na at 850 MeV,
as functions of the scattering angle $\theta$.}
\label{figO}
\end{figure}

\begin{figure}
\caption{Parity-violating asymmetry parameters $A_l$
(upper panel) and squares of normalized Fourier transforms
of neutron densities (lower panel), as functions of the
momentum transfer $q$, for elastic scattering from
$^{23-27}$Na at 850 MeV.}
\label{figP}
\end{figure}

\begin{figure}
\caption{Same as in Fig. \ref{figP}, but for 
elastic scattering from $^{28-32}$Na at 850 MeV.}
\label{figQ}
\end{figure}

\begin{figure}
\caption{Parity-violating asymmetry parameters $A_l$
(upper panel) and squares of normalized Fourier transforms
of neutron densities (lower panel), as functions of the
momentum transfer $q$, for elastic scattering from
$^{30,32,34}$Ne at 850 MeV.}
\label{figR}
\end{figure}


\begin{references}
\bibitem{Bat.89} C. J. Batty, E. Friedman, H. J. Gils, 
	and H. Rebel, Adv. Nucl. Phys. {\bf 19}, 1 (1989).
\bibitem{DDS.89} T.W. Donnelly, J. Dubach, and Ingo Sick,
	Nucl. Phys. {\bf A503}, 589 (1989).
\bibitem{Hor.98} C. J. Horowitz, Phys. Rev. C {\bf 57}, 3430 (1998).
\bibitem{PVL.97} W. P\"oschl, D. Vretenar, G.A. Lalazissis,
	and P. Ring, Phys. Rev. Lett. {\bf 79}, 3841 (1997).
\bibitem{LVP.98} G.A. Lalazissis, D. Vretenar, W. P\"oschl,
	and P. Ring, Nucl. Phys. {\bf A632}, 363 (1998).
\bibitem{LVR.98} G.A. Lalazissis, D. Vretenar,
	and P. Ring, Phys. Rev. C {\bf 57}, 2294 (1998).
\bibitem{LVR.98a}  G.A. Lalazissis, D. Vretenar, P. Ring, M. Stoitsov, and
L. Robledo,  Phys. Rev. C {\bf 60}, 014310 (1999).
\bibitem{LVR.99} G.A. Lalazissis, D. Vretenar,
	and P. Ring, Nucl. Phys. {\bf A650}, 133 (1999).
\bibitem{VLR.99} D. Vretenar,  G.A. Lalazissis, and P. Ring,
	Phys. Rev. Lett. {\bf 82}, 4595 (1999).
\bibitem{LVR.99a} G.A. Lalazissis, D. Vretenar,
	and P. Ring, Phys. Rev. C {\bf 60}, 051302 (1999).
\bibitem{DNW.96} J. Dobaczewski, W. Nazarewicz, T. R. Werner,
	J. F. Berger, C. R. Chinn, and J.Decharg\' e,
		Phys. Rev. C {\bf 53}, 2809 (1996).
\bibitem{Rin.96} P. Ring, Progr. Part. Nucl. Phys. {\bf 37}, 193 (1996).
\bibitem{SW.97}B.D. Serot and J.D. Walecka,
	 Adv. Nucl. Phys. {\bf 16}, 1 (1986);
		 Int. J. Mod. Phys. {\bf E6}, 515 (1997).
\bibitem{LKR.97}  G. A. Lalazissis, J. K\"{o}nig and P. Ring; Phys. Rev.
	{\bf C55}, 540 (1997).
\bibitem{BGG.84}  J. F. Berger, M. Girod and D. Gogny; Nucl. Phys.
		 {\bf A428}, 32 (1984).
\bibitem{Kra.99} A. Krasznahorkay  {\it et al.}, Phys. Rev. Lett. {\bf 82}, 
	3216 (1999).
\bibitem{Suz.95} T. Suzuki {\it et al.}, Phys. Rev. Lett. {\bf 75},
	3241 (1995).
\bibitem{Tani.85} I. Tanihata et al., Phys. Rev. Lett. {\bf 55},
	2676 (1985); Phys. Lett. {\bf B206}, 592 (1988).
\bibitem{Tani.95} I. Tanihata, Prog. Part. Nucl. Phys.
	{\bf 35}, 505 (1995).
\bibitem{HJJ.95} P. Hansen, A.S. Jensen, and B. Jonson,
		 Annu. Rev. Nucl. Part. Phys. {\bf 45}, 591 (1995).
\bibitem{MR.96} J. Meng and P. Ring,
		Phys. Rev. Lett. {\bf 77}, 3963 (1996).
\bibitem{Fro.77} B. Frois {\it et al}, Phys. Rev. Lett. {\bf 38}, 152 (1977).
\end{references}
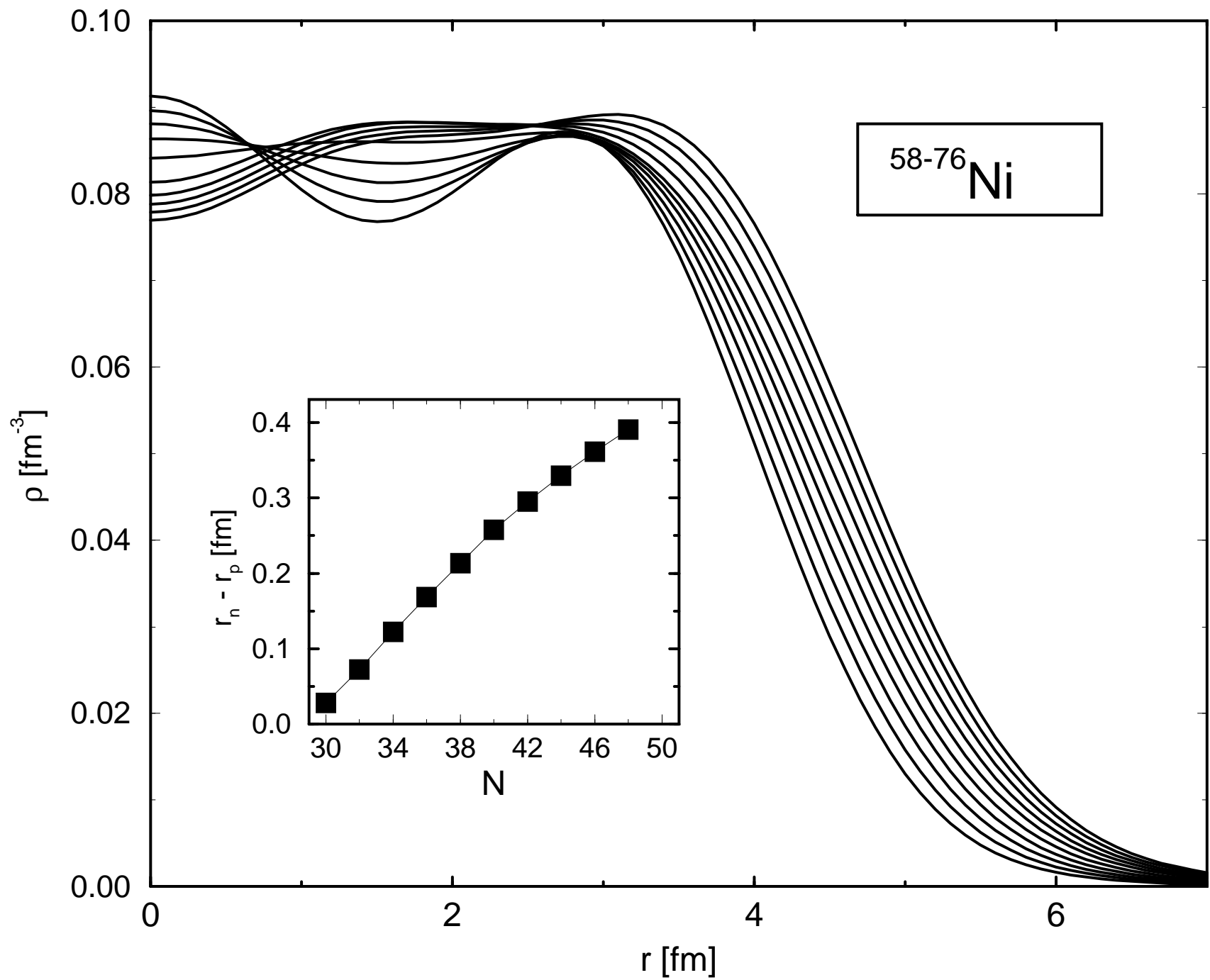
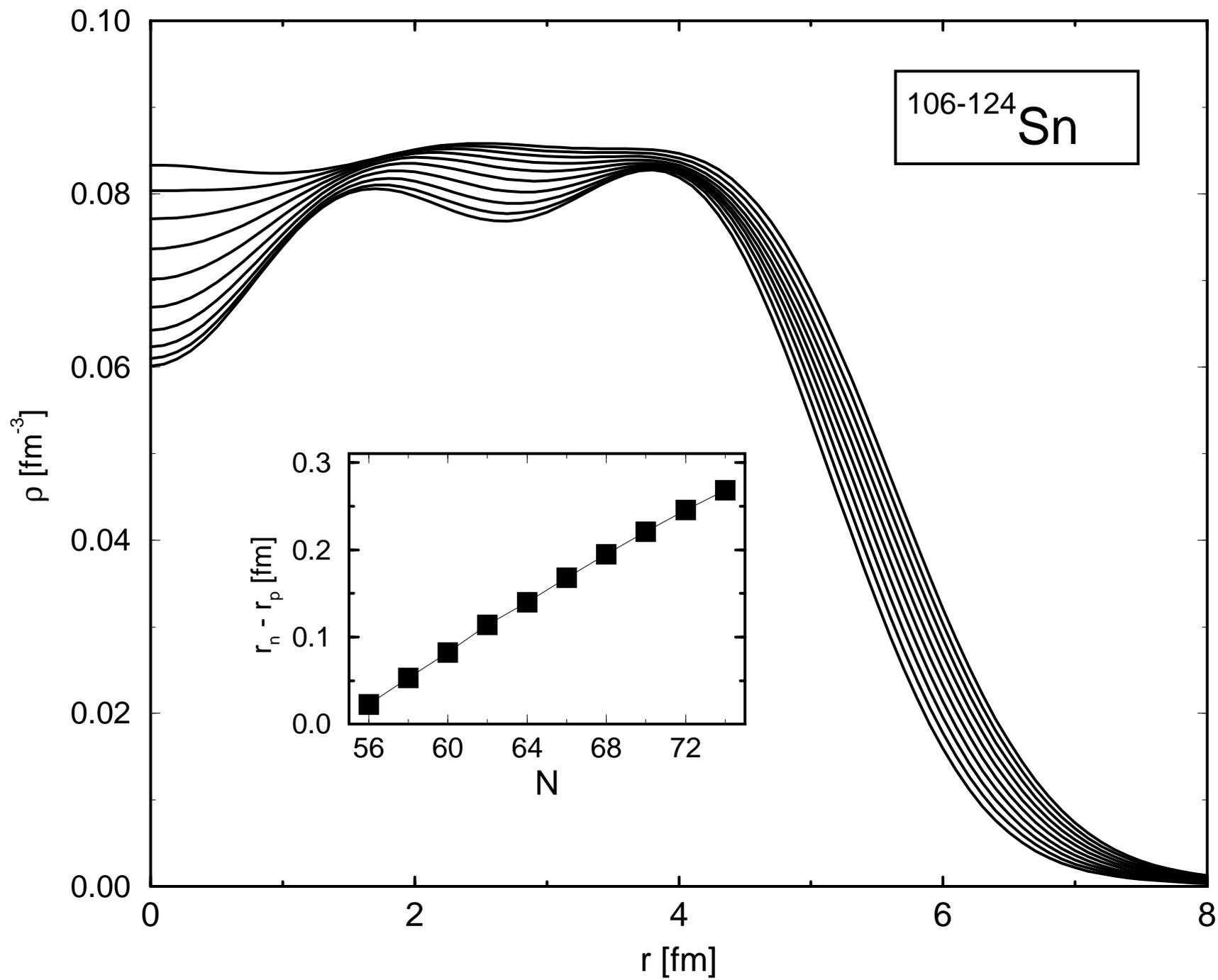
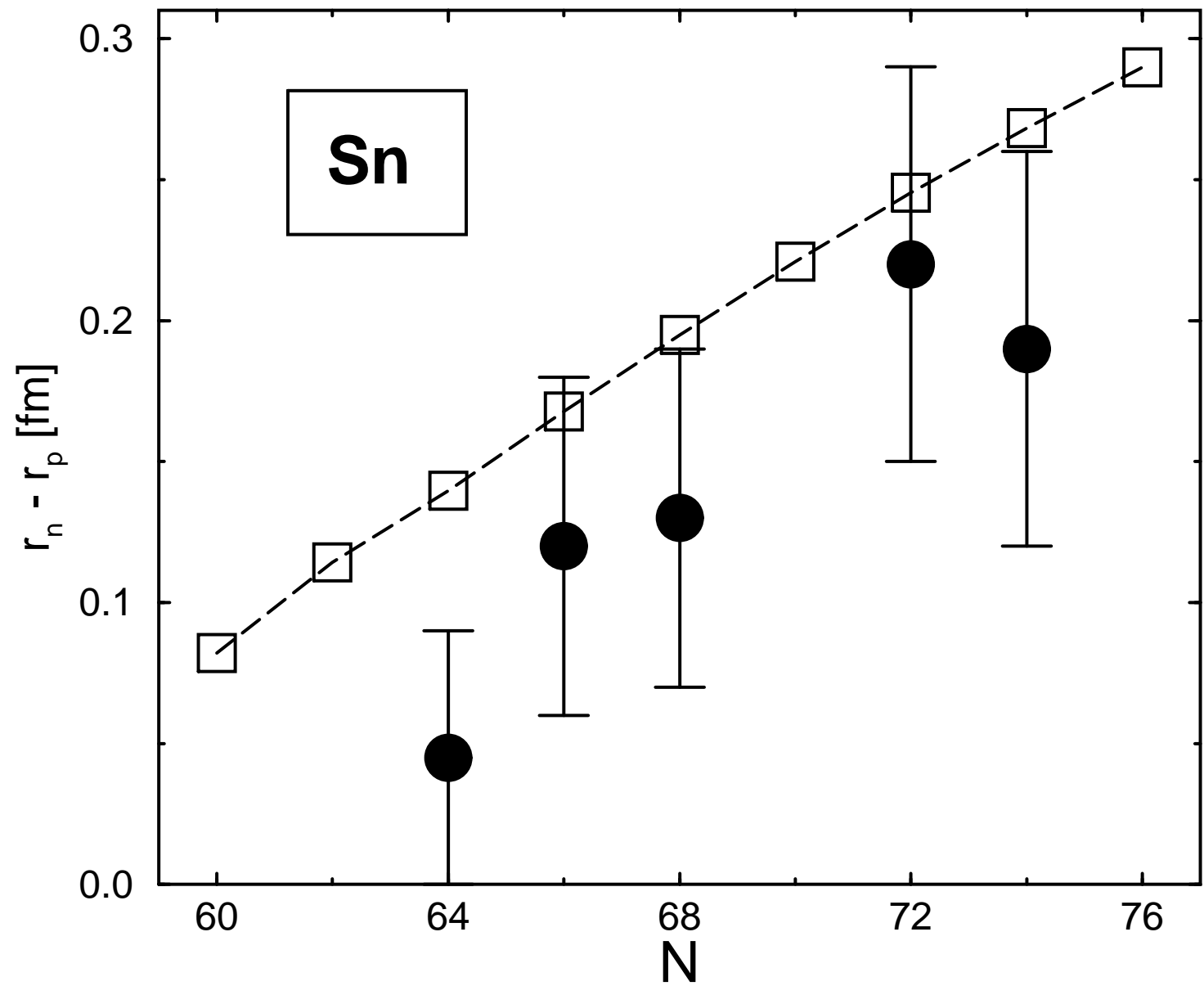
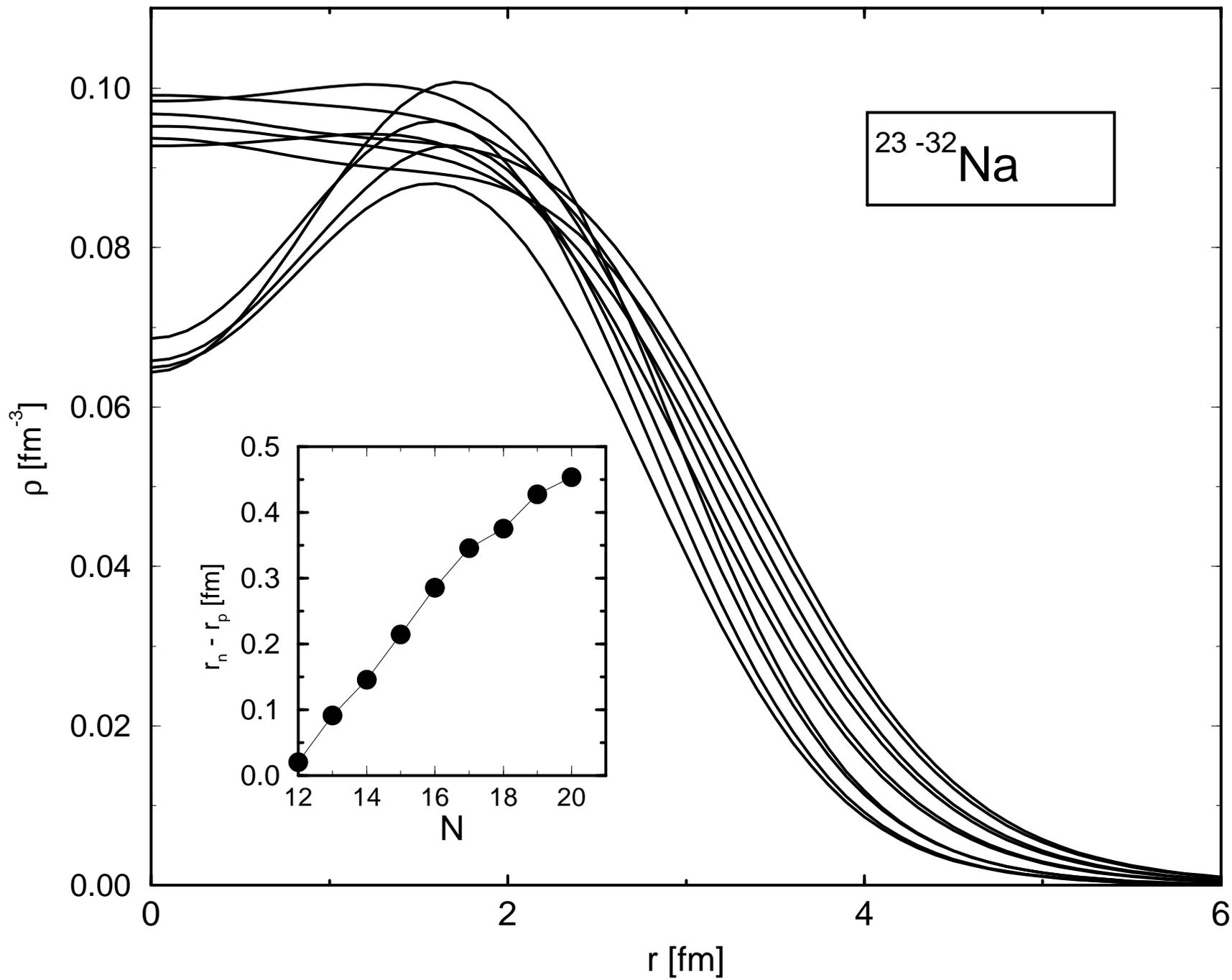
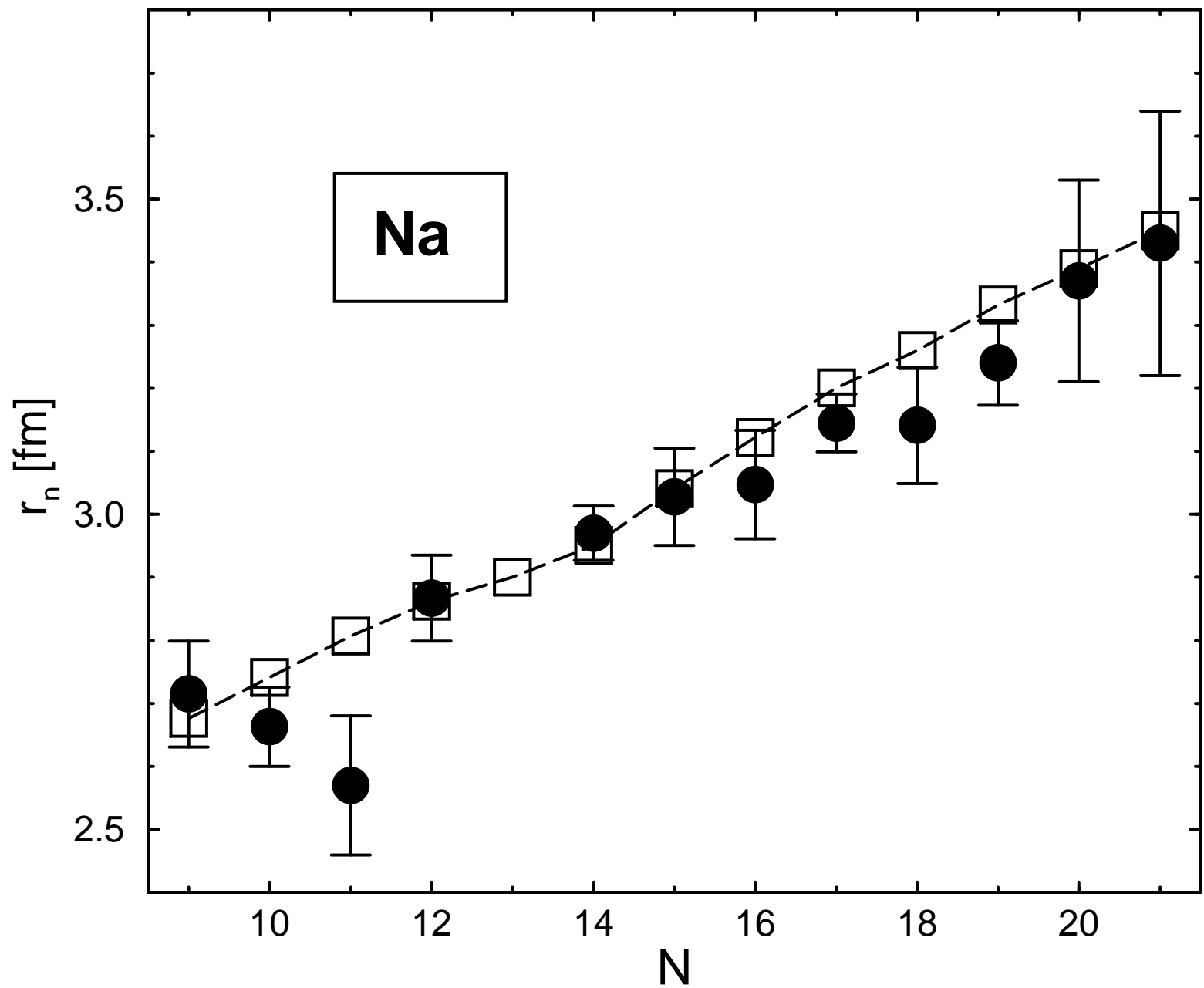
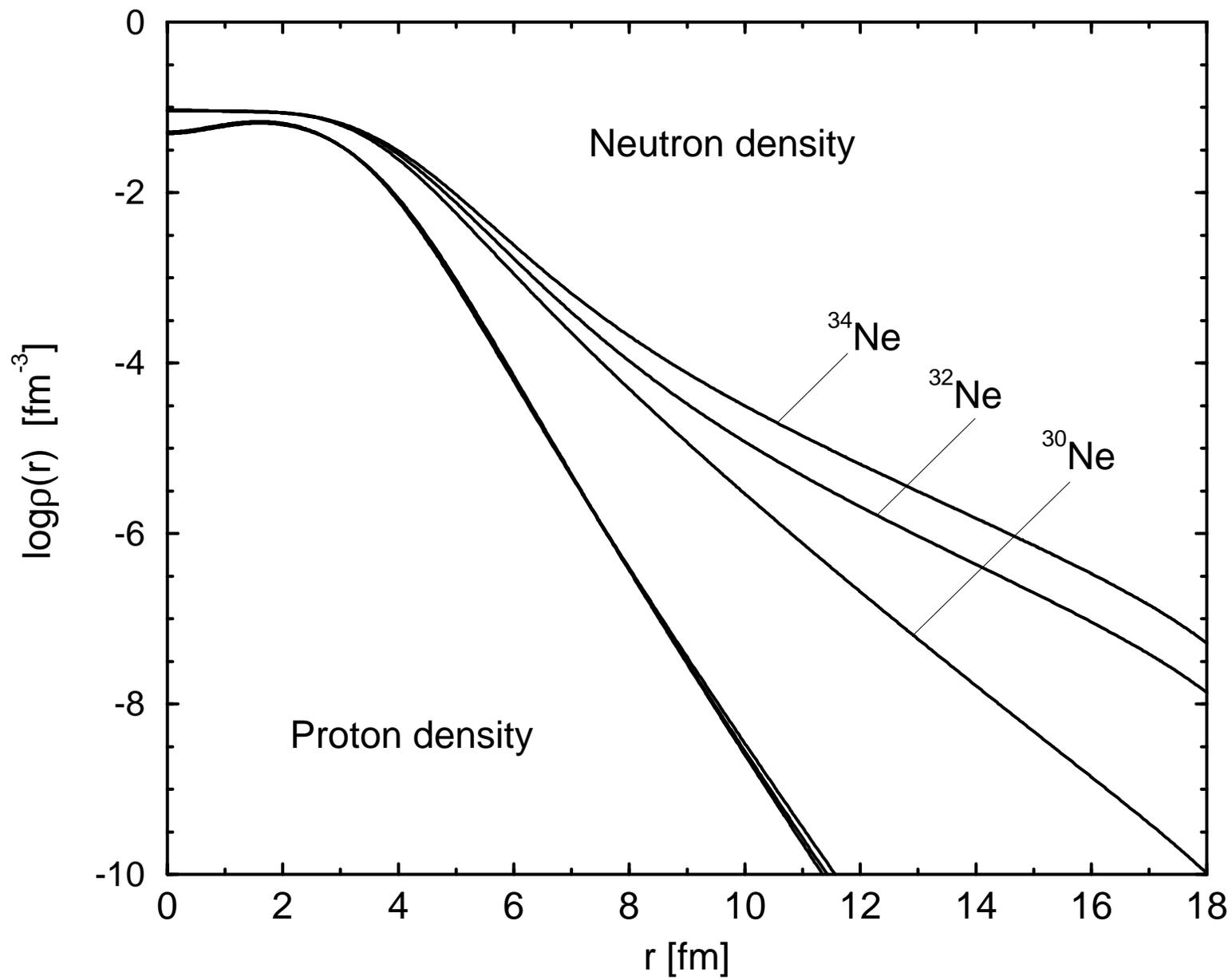
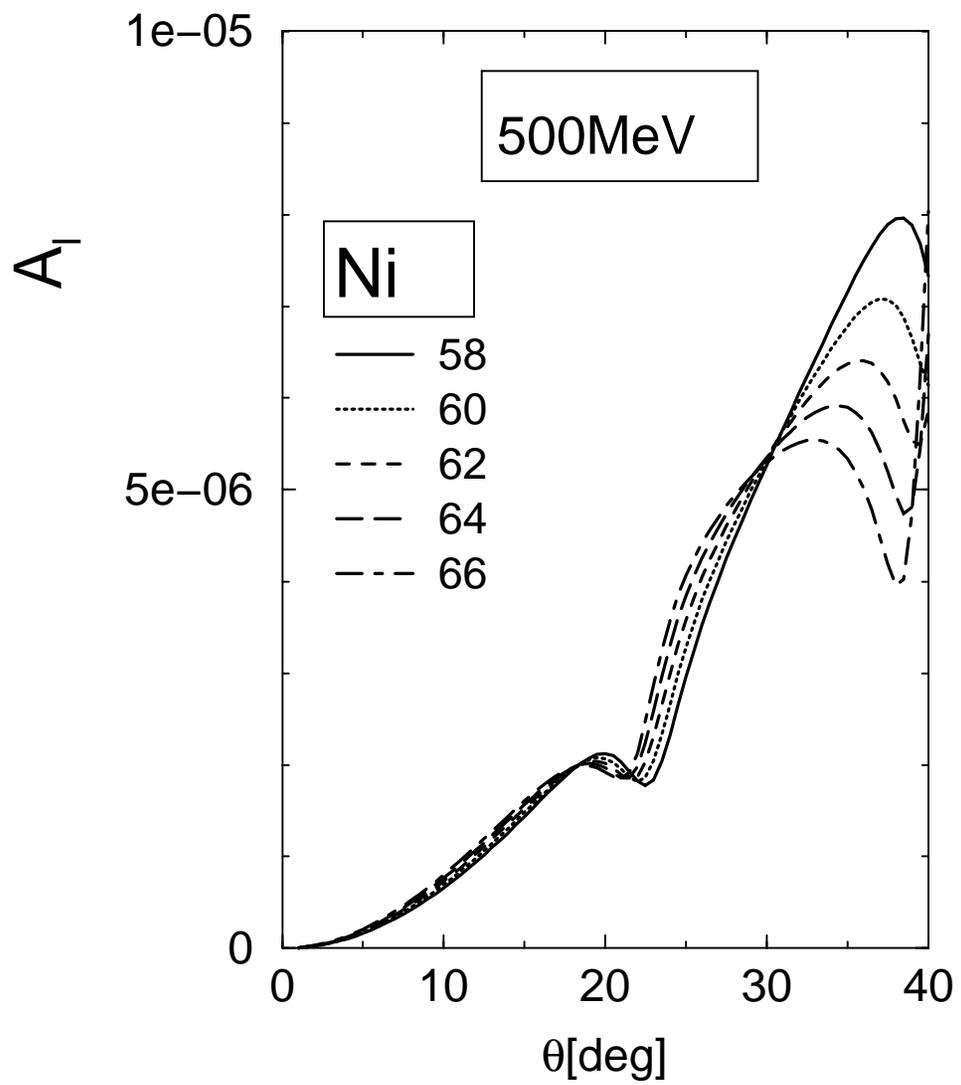
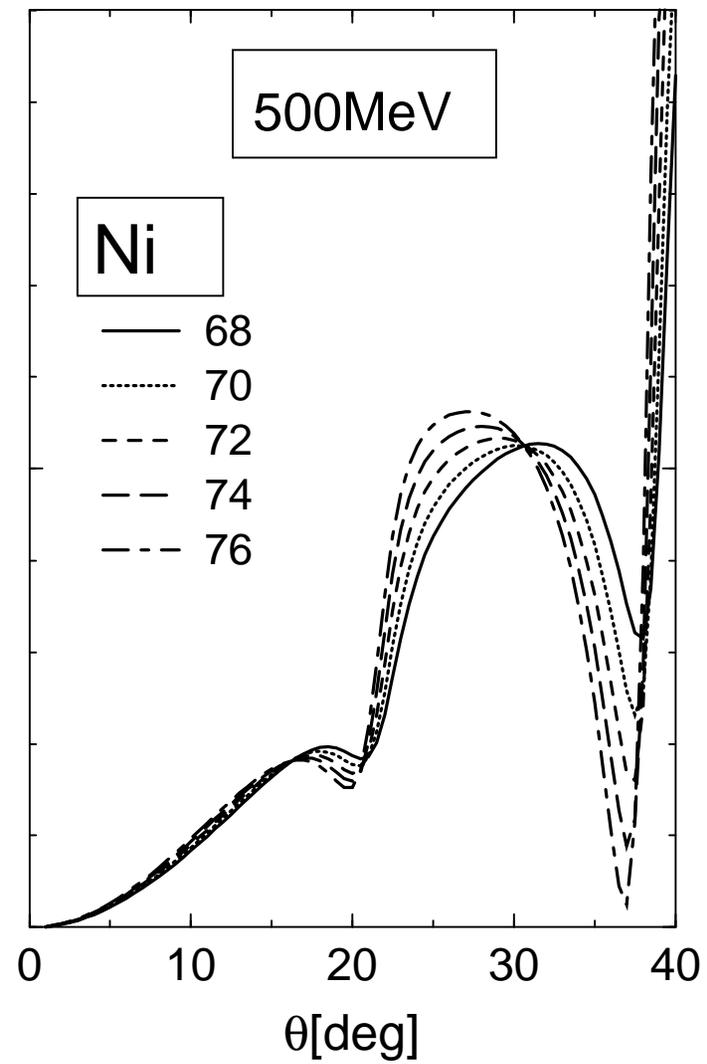
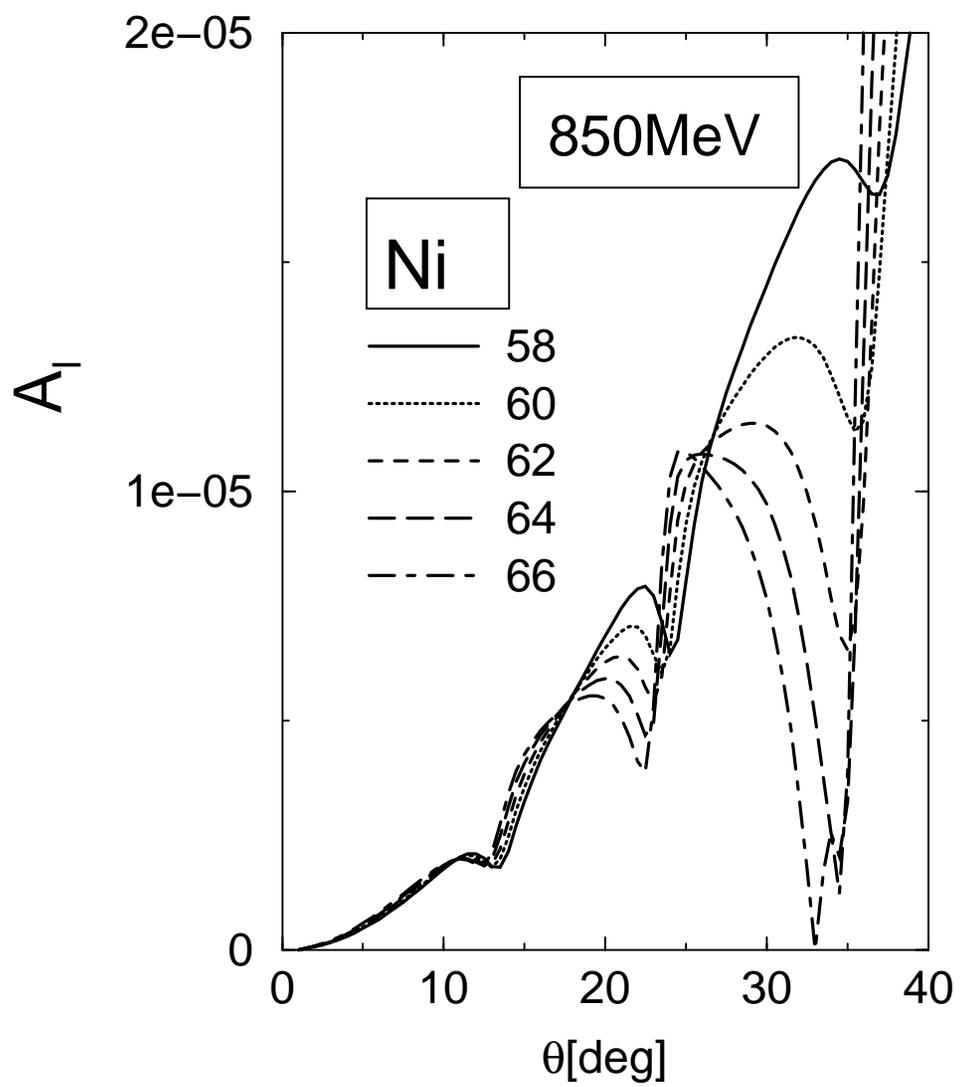
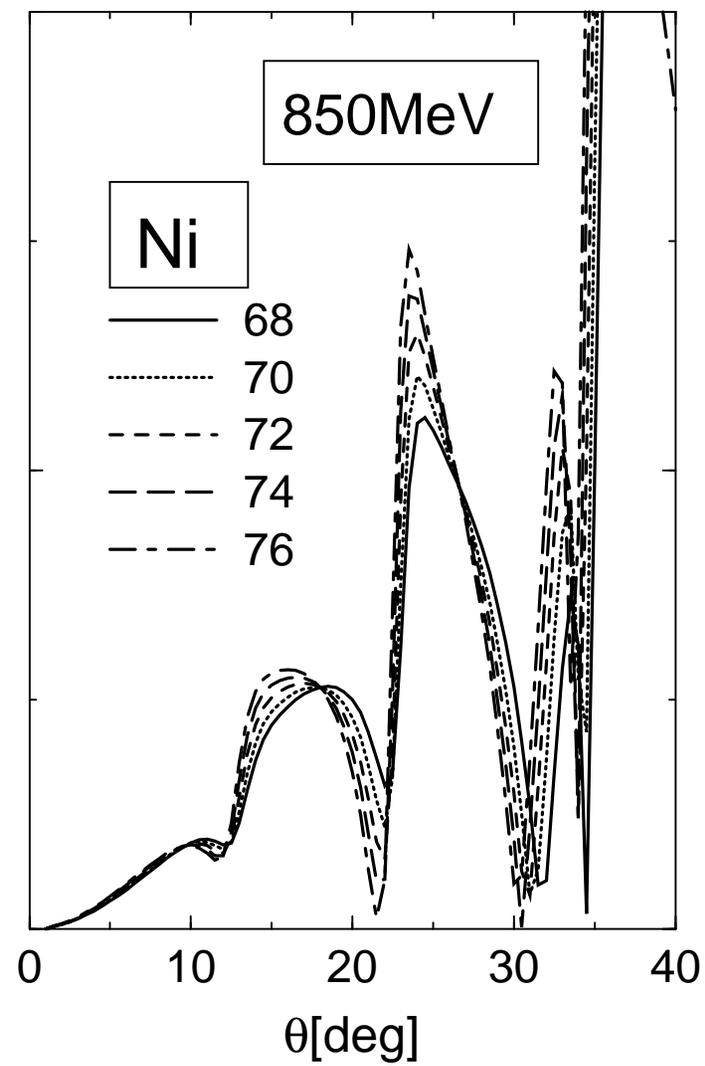
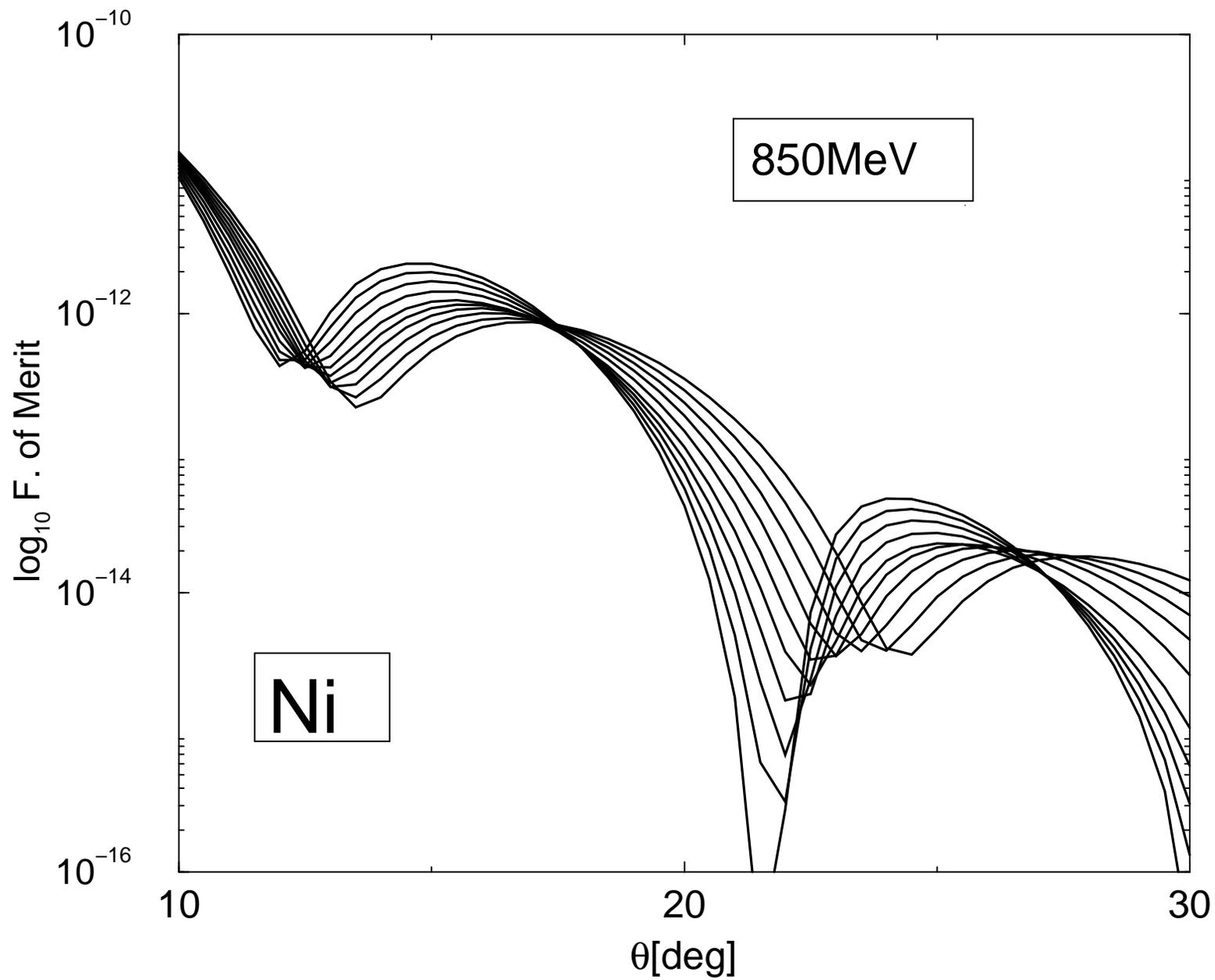
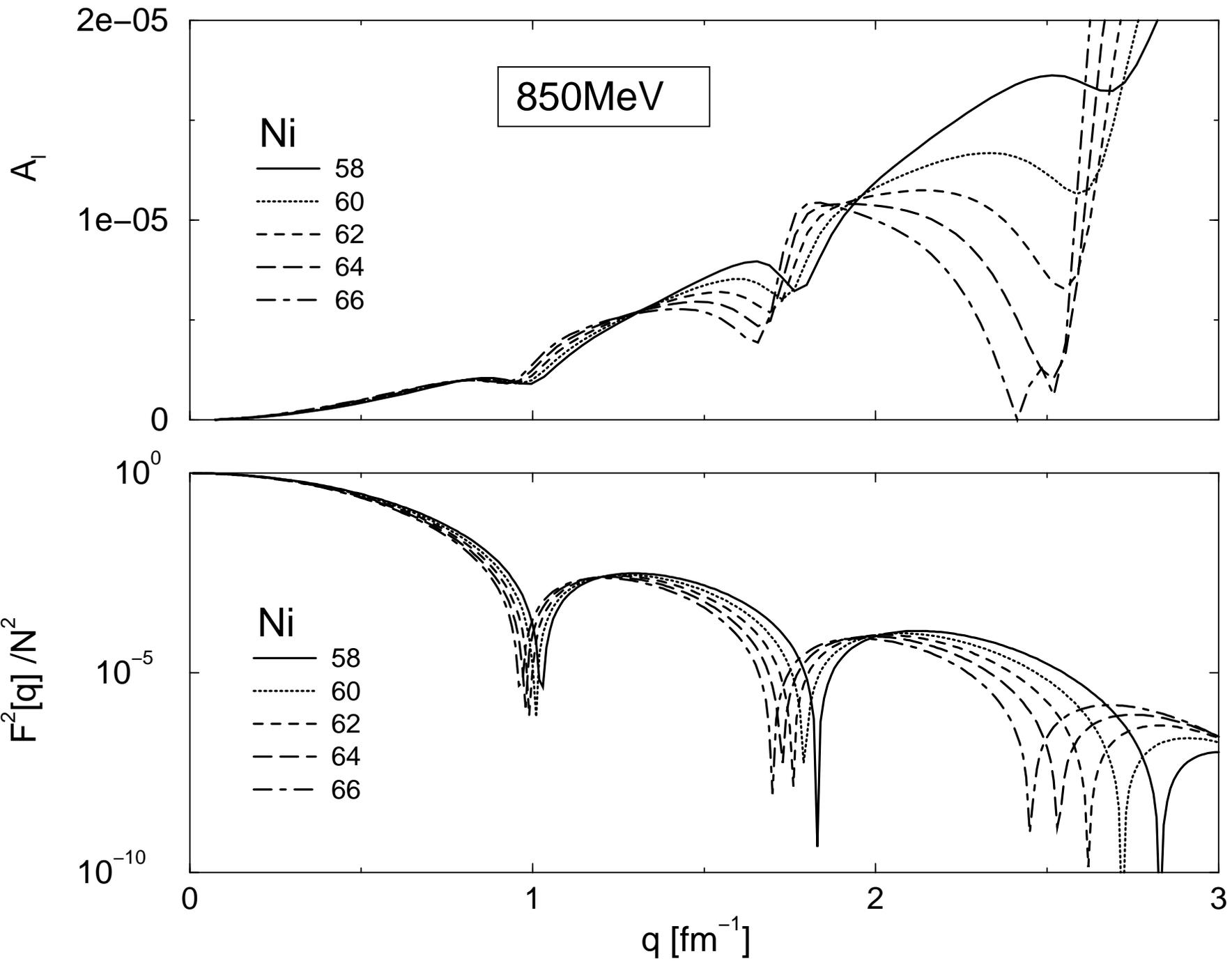
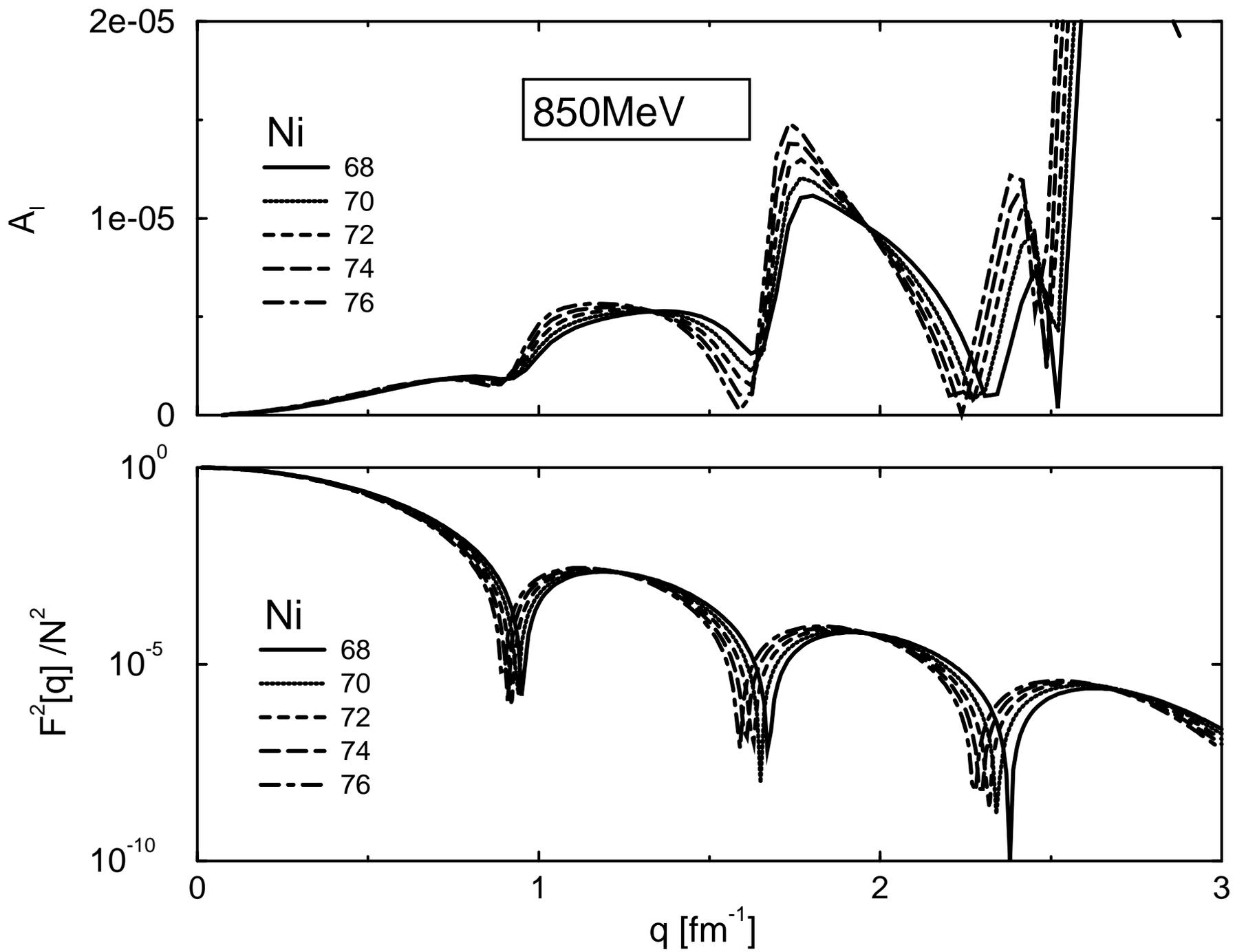
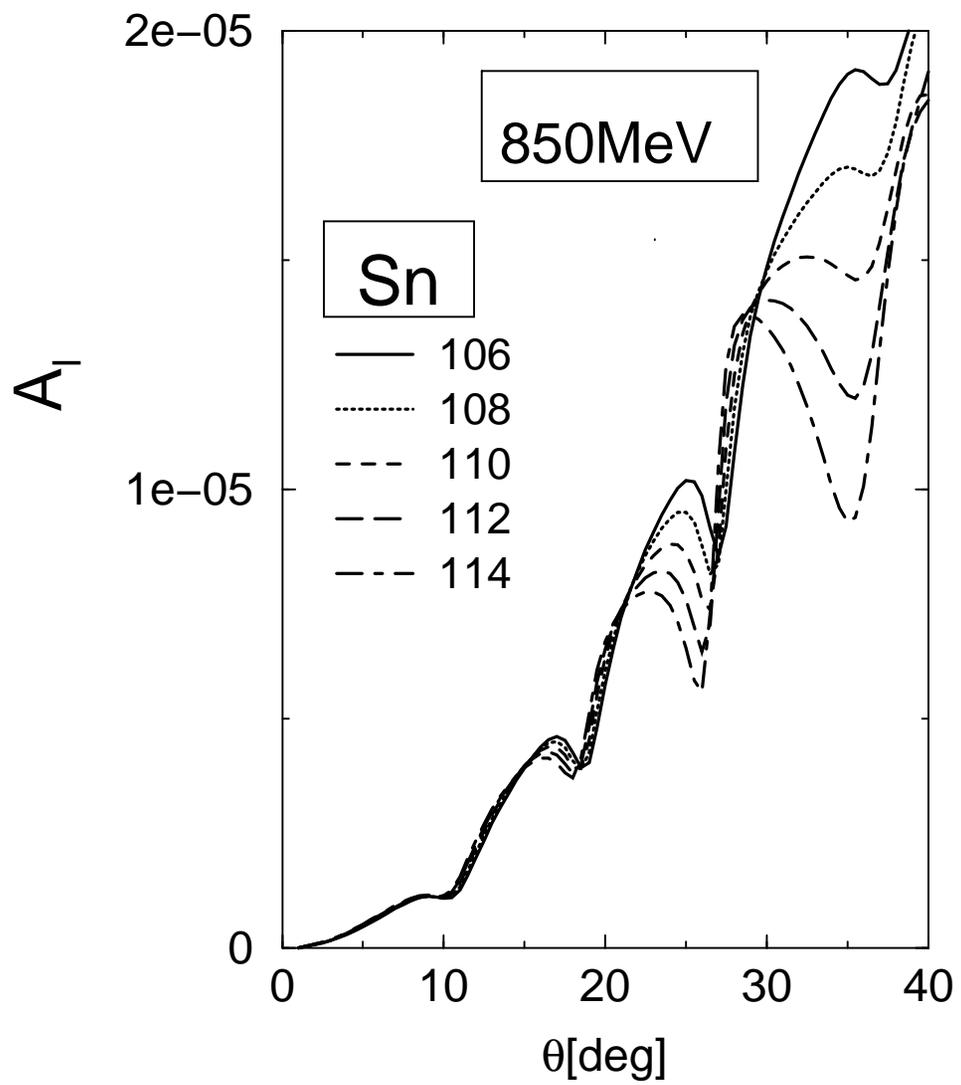
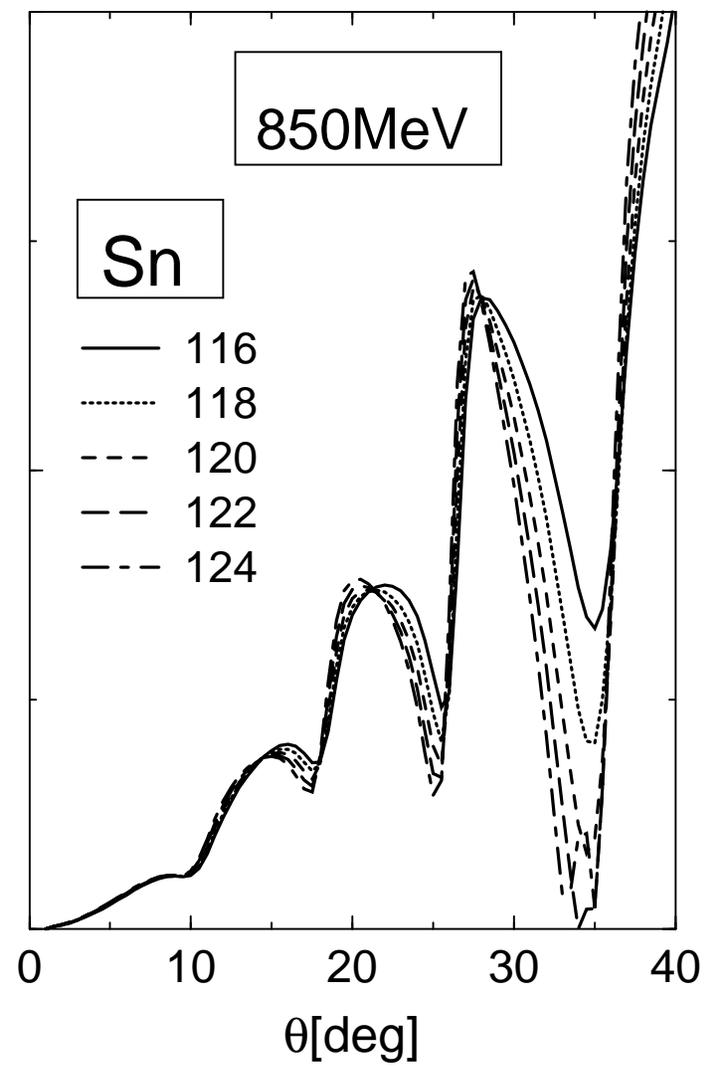
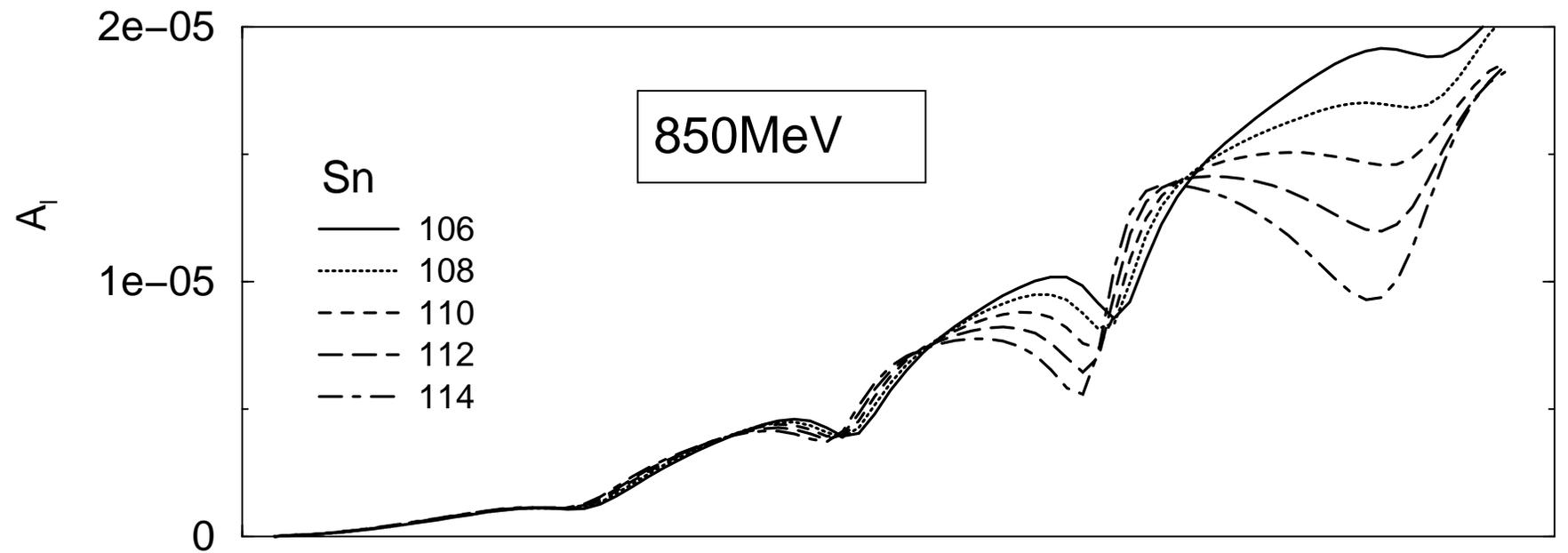
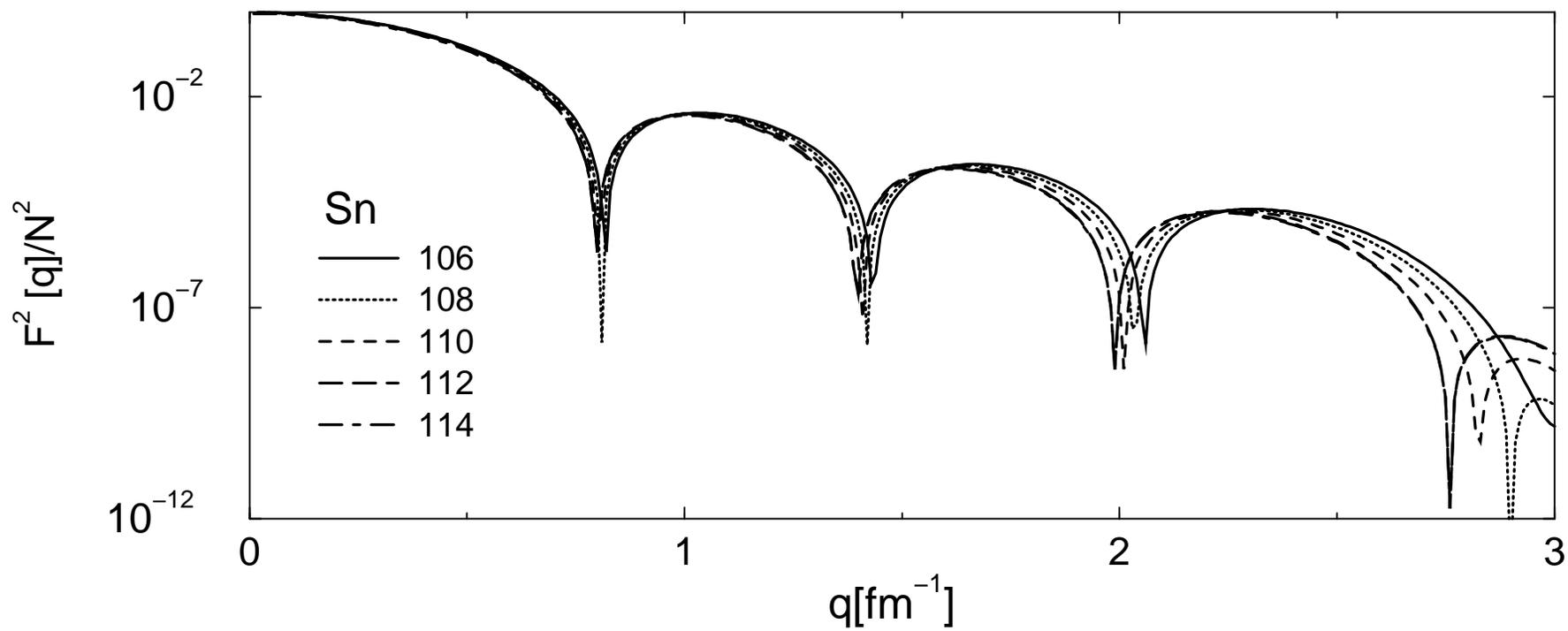
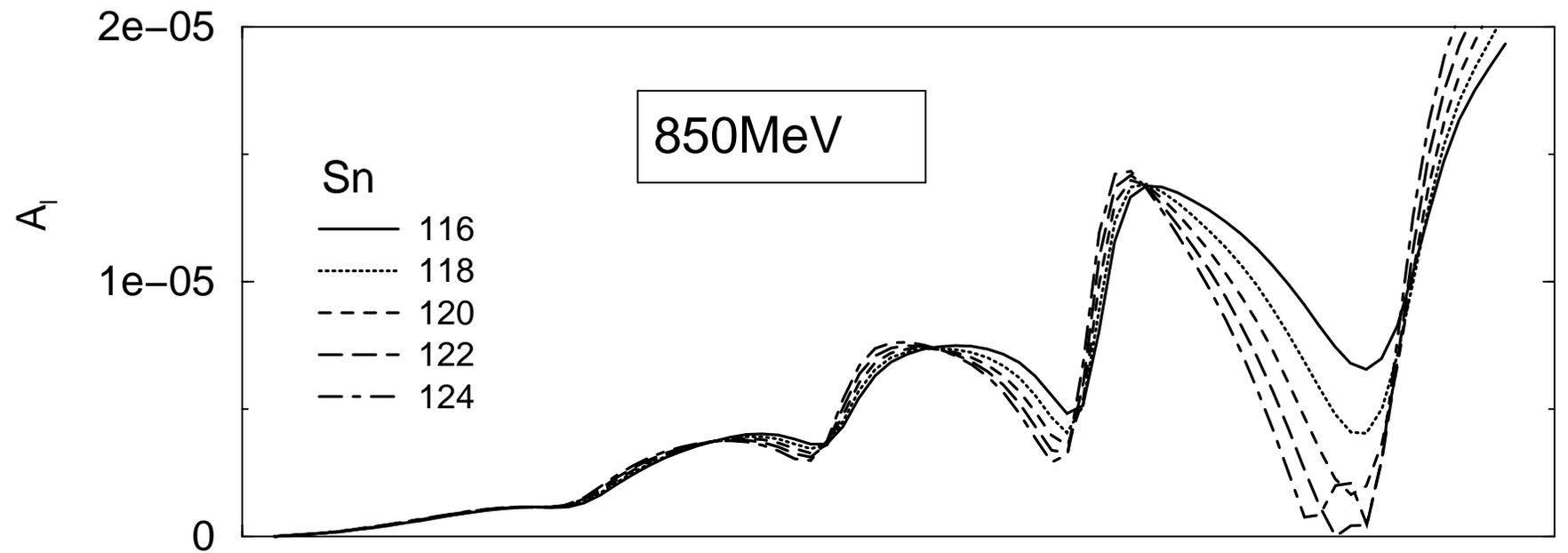
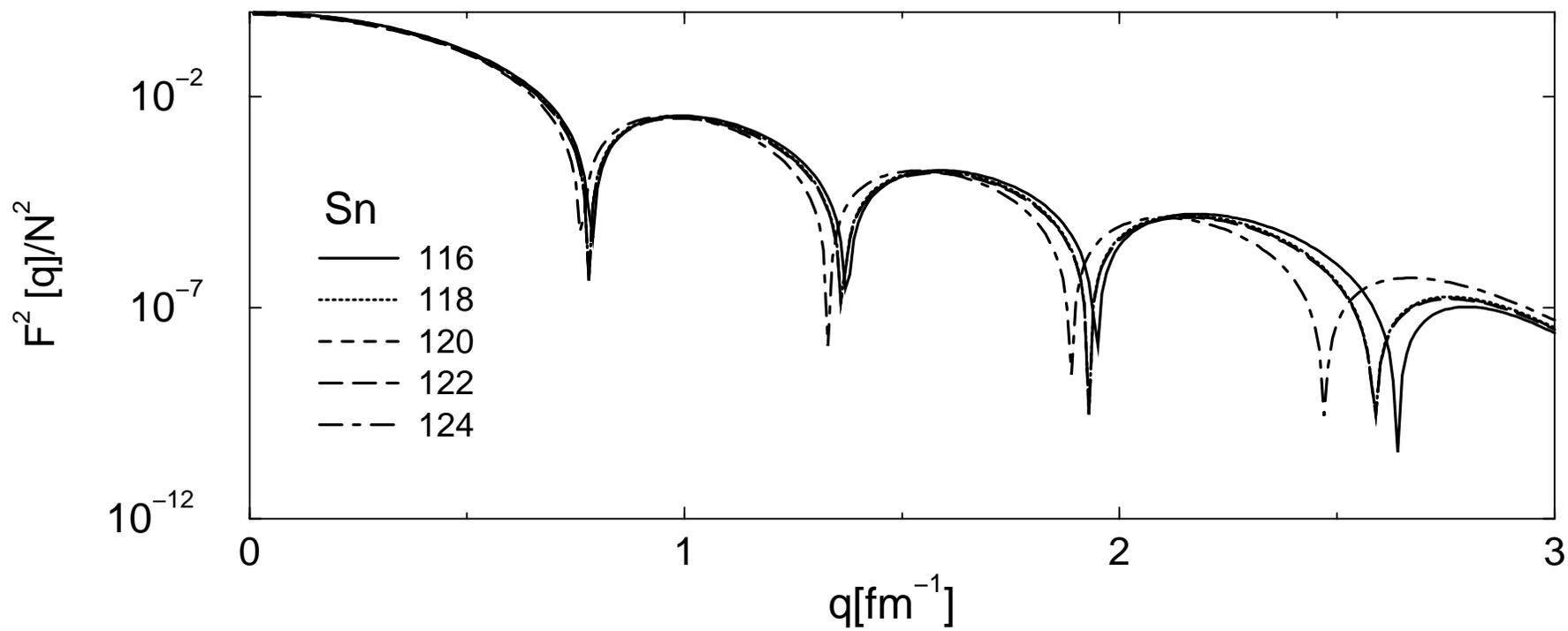
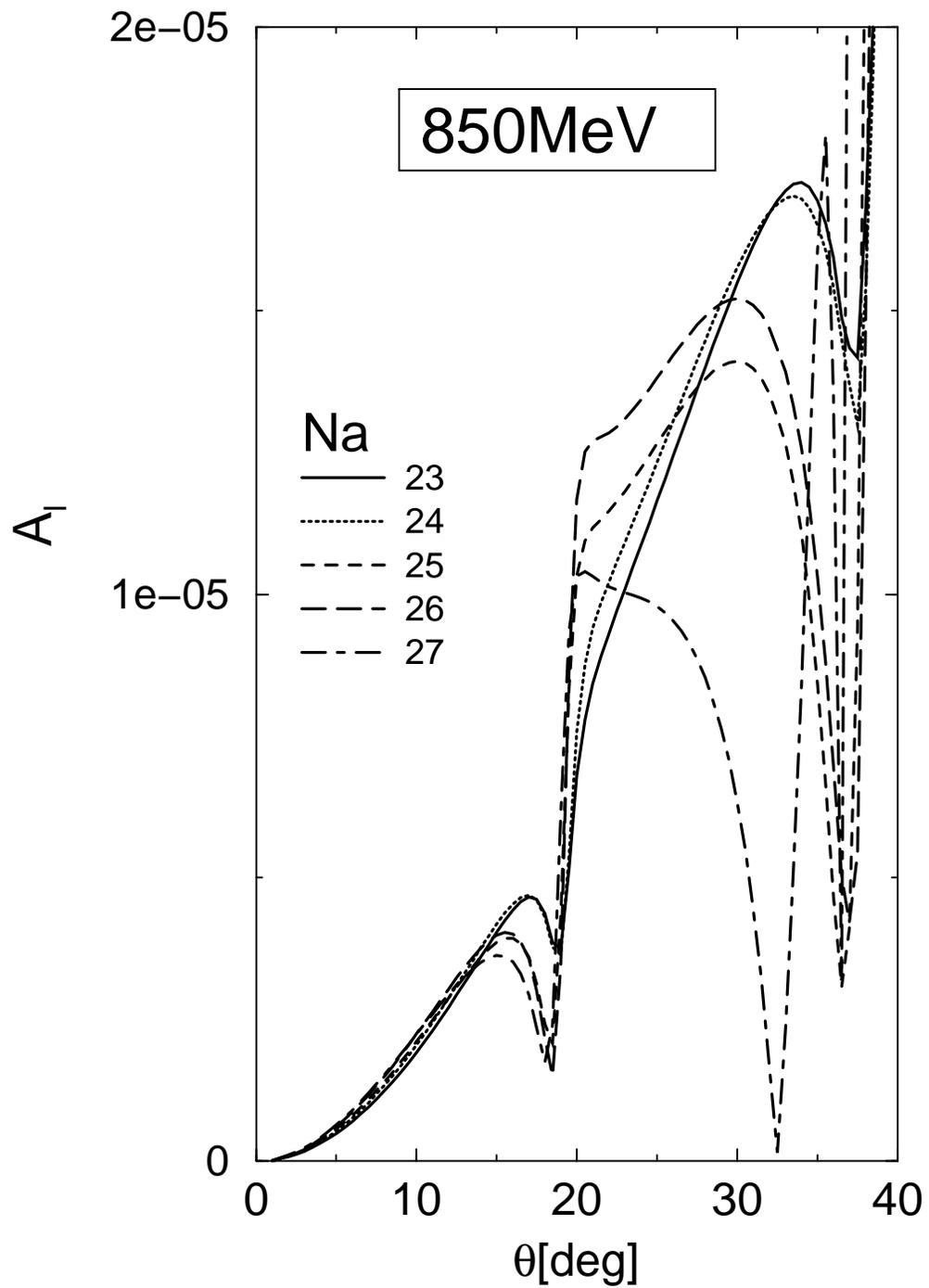
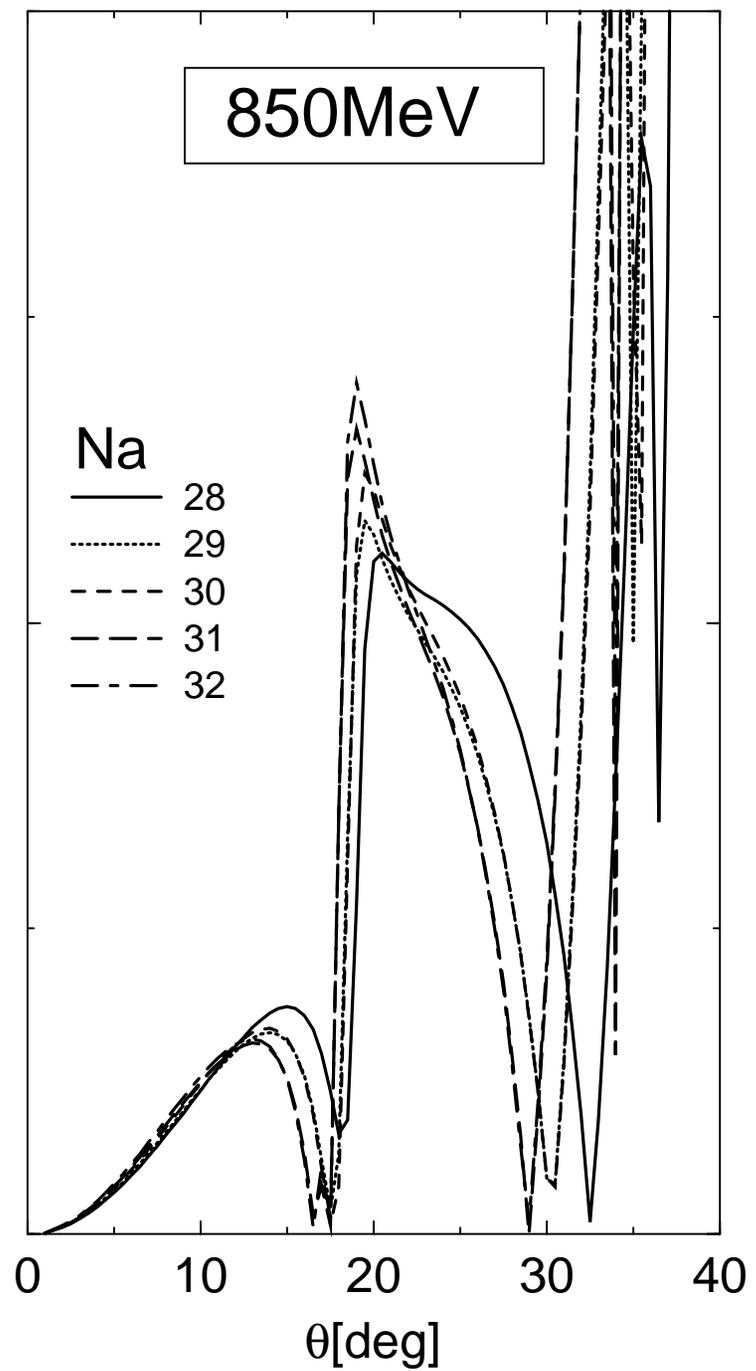
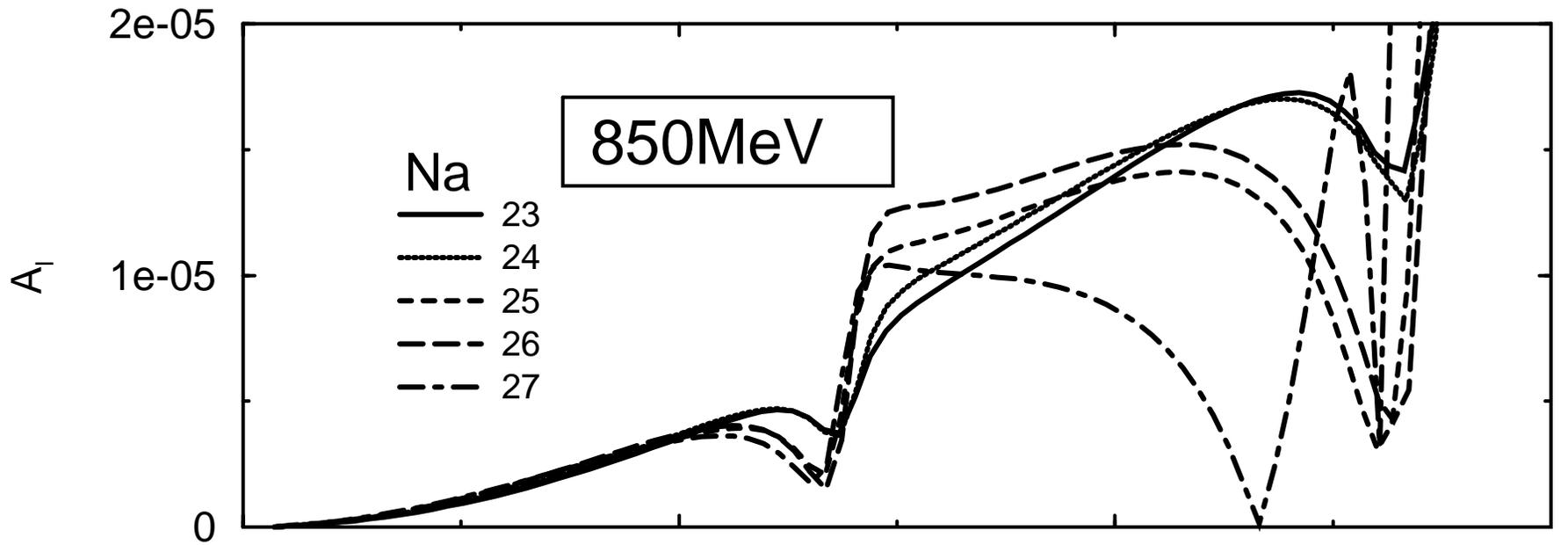
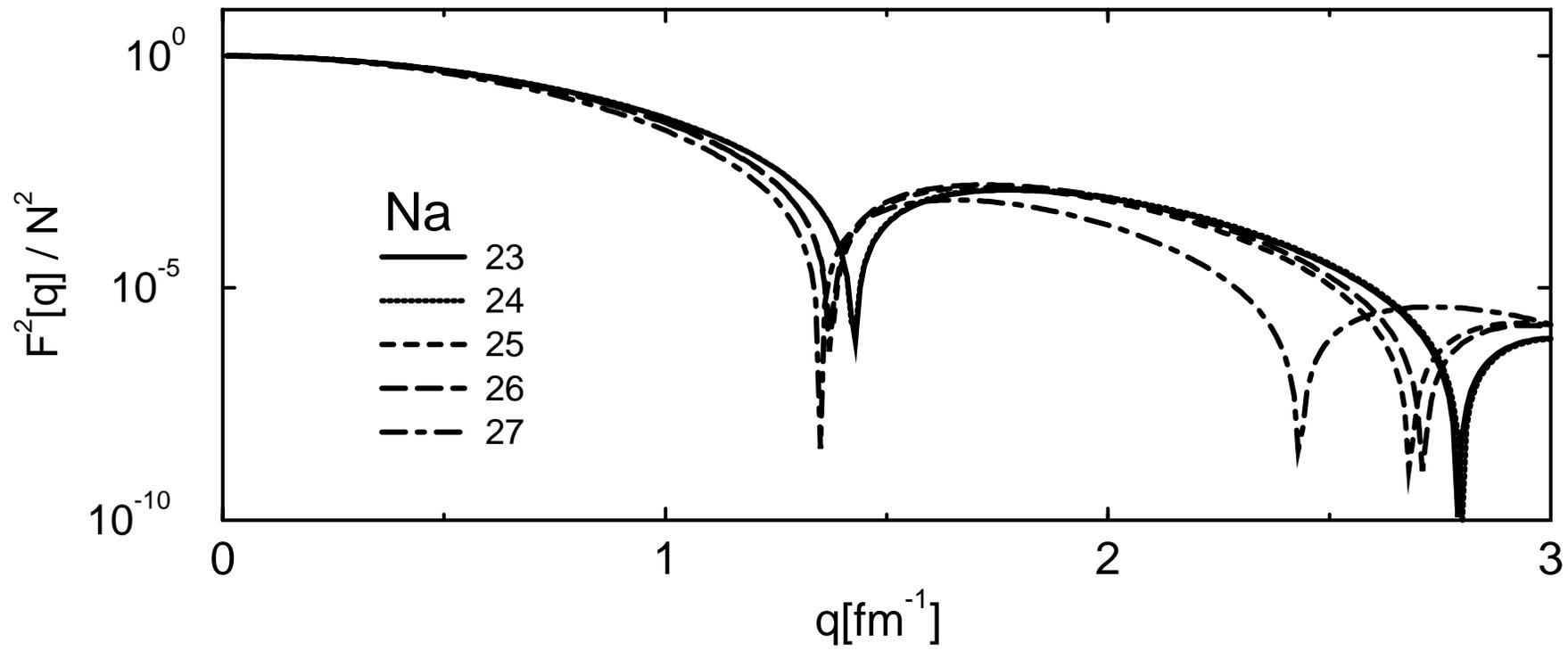
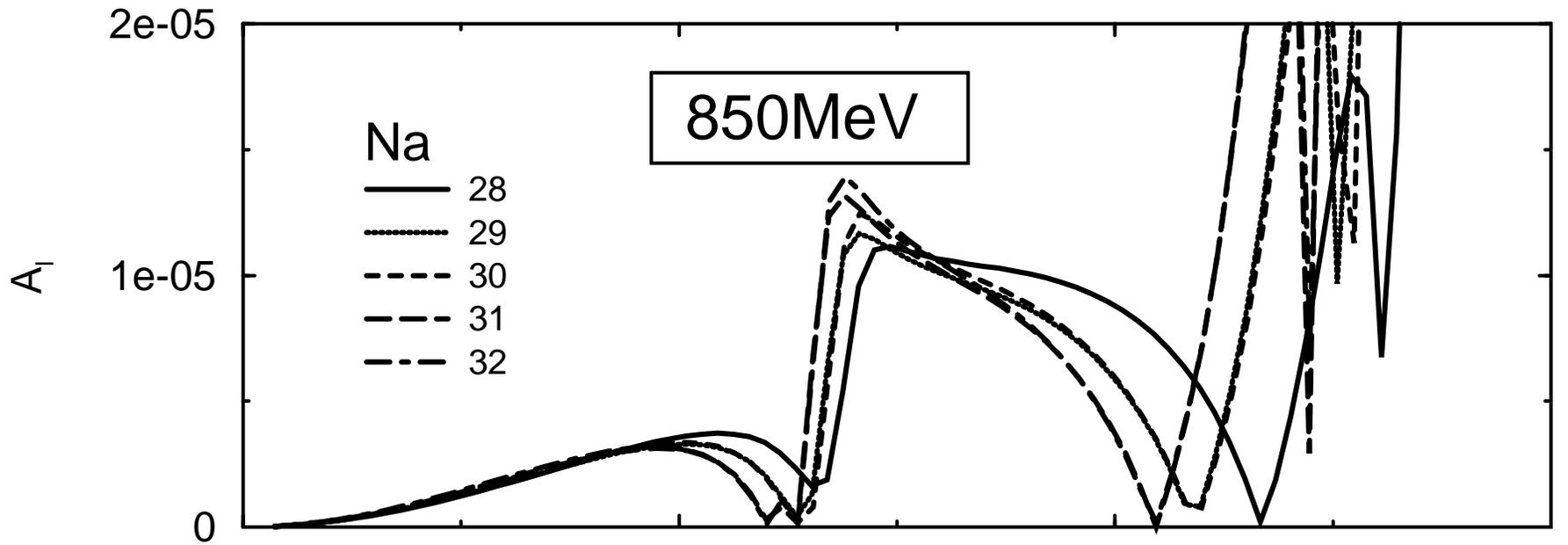
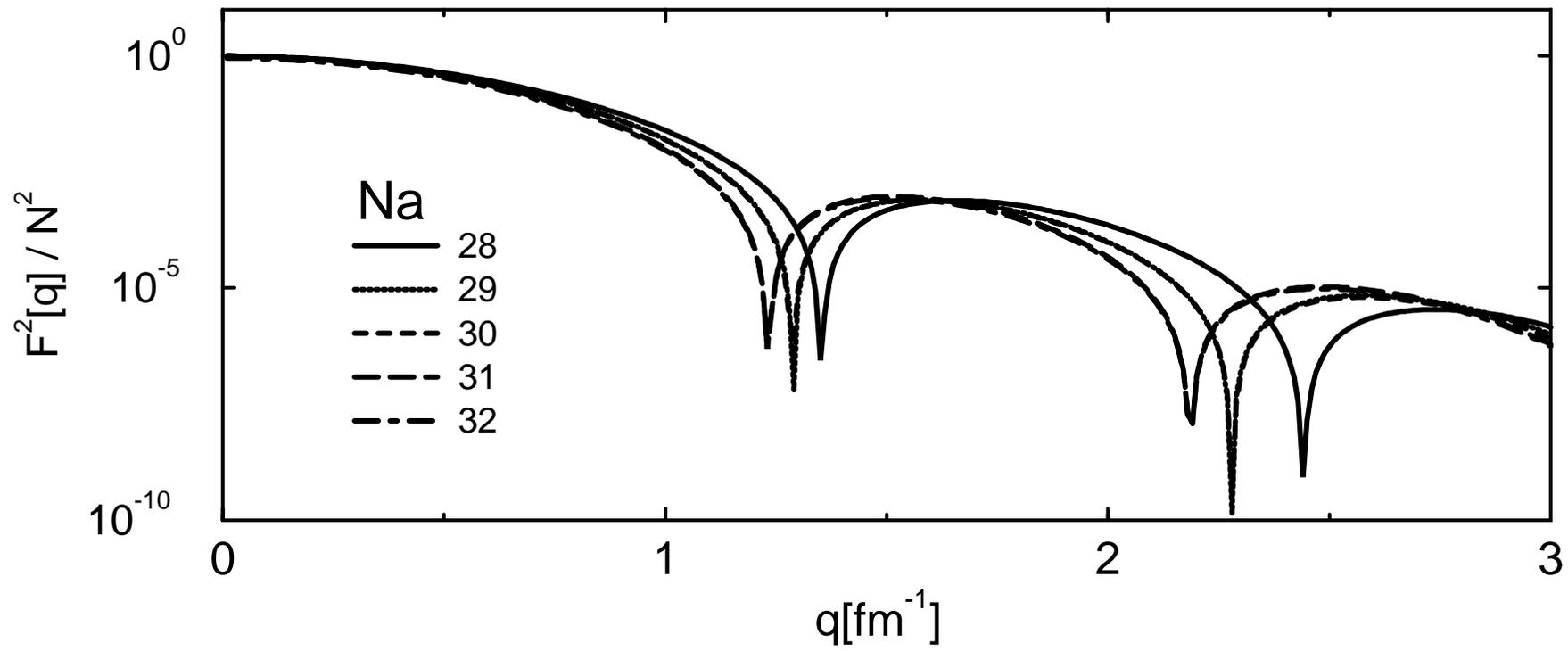
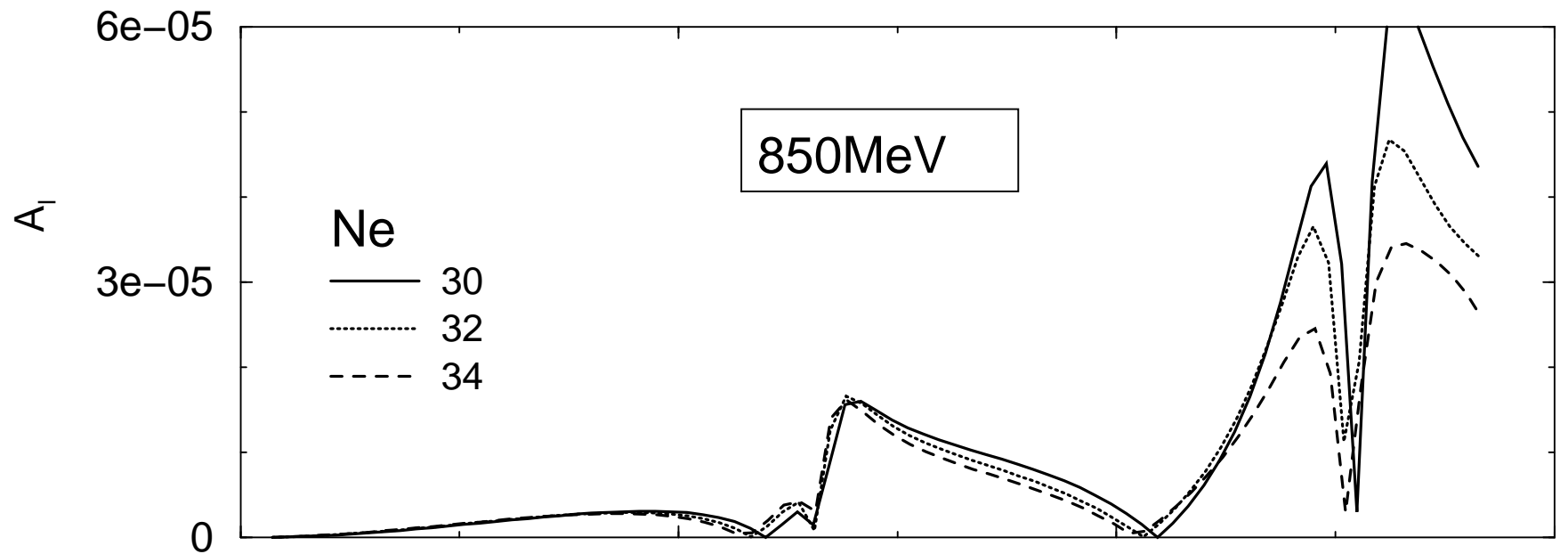
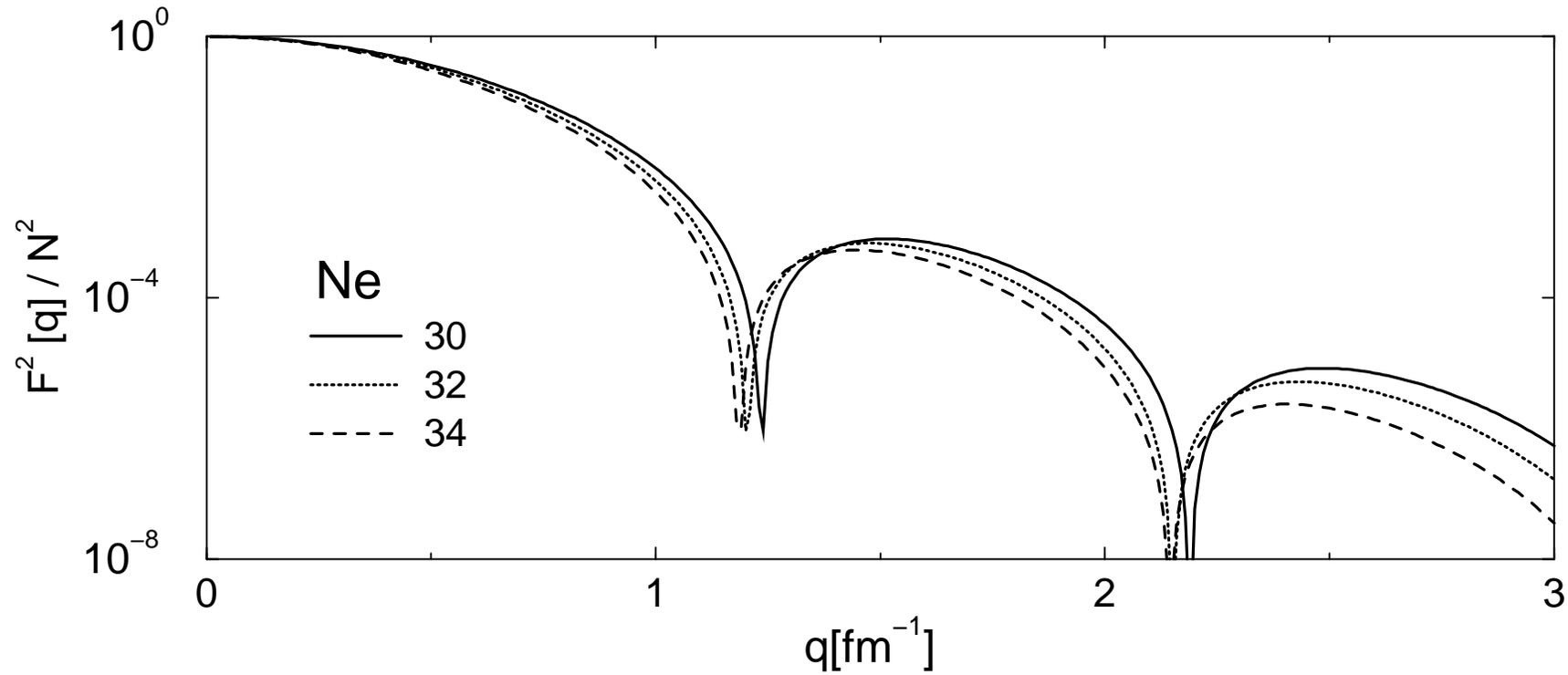
\end{document}